\documentclass[pra,twocolumn,amsmath,amssymb,floatfix,reprint,footinbib,superscriptaddress,longbibliography,showkeys]{revtex4-1}
\usepackage{dcolumn}
\usepackage{graphicx}
\usepackage{mathrsfs}
\usepackage{mdwlist}
\usepackage{subfigure}
\usepackage{booktabs}
\usepackage{amsmath}
\usepackage{extarrows}
\usepackage{dsfont}
\usepackage{amstext}
\usepackage{amssymb}
\usepackage{amsbsy}
\usepackage{bbm}
\usepackage{bm}
\usepackage{cases}
\usepackage{amsthm}
\usepackage{graphicx}
\usepackage{textcomp}
\usepackage{multirow}
\usepackage{color}
\usepackage{diagbox}
\usepackage[colorlinks,citecolor=blue]{hyperref}
\definecolor{Dgreen}{RGB}{0, 100, 0}
\usepackage{url}
\usepackage[colorlinks]{hyperref}
\hypersetup{%
	plainpages=true,
	breaklinks=true,  
	hypertexnames=false,  
	pageanchor=true,
	colorlinks=true,
	linkcolor={blue},
	citecolor={blue},
	urlcolor={blue},
	anchorcolor={black}
}

\begin{document}
\title{Robust population inversion in three-level systems by composite pulses}
\date{\today}

\author{Cheng Zhang}
\affiliation{Fujian Key Laboratory of Quantum Information and Quantum Optics (Fuzhou University), Fuzhou
350108, China}
\affiliation{Department of Physics, Fuzhou University, Fuzhou 350108, China}

\author{Yang Liu}
\affiliation{Fujian Key Laboratory of Quantum Information and Quantum Optics (Fuzhou University), Fuzhou
350108, China}
\affiliation{Department of Physics, Fuzhou University, Fuzhou 350108, China}

\author{Zhi-Cheng Shi}\thanks{szc2014@yeah.net}
\affiliation{Fujian Key Laboratory of Quantum Information and Quantum Optics (Fuzhou University), Fuzhou
350108, China}
\affiliation{Department of Physics, Fuzhou University, Fuzhou 350108, China}

\author{Jie Song}
\affiliation{Department of Physics, Harbin Institute of Technology, Harbin 150001, China}

\author{Yan Xia}\thanks{xia-208@163.com}
\affiliation{Fujian Key Laboratory of Quantum Information and Quantum Optics (Fuzhou University), Fuzhou
350108, China}
\affiliation{Department of Physics, Fuzhou University, Fuzhou 350108, China}

\author{Shi-Biao Zheng}
\affiliation{Fujian Key Laboratory of Quantum Information and Quantum Optics (Fuzhou University), Fuzhou
350108, China}
\affiliation{Department of Physics, Fuzhou University, Fuzhou 350108, China}

\begin{abstract}
In this work, we exploit the idea of composite pulses to achieve robust population inversion in a three-level quantum system.
The scheme is based on the modulation of the coupling strength, while the other physical parameters remain unchanged. The composite pulses sequence is designed by vanishing high-order error terms, and can compensate the systematic errors to any desired order.
In particular, this scheme keeps a good performance under the disturbance of waveform deformations.
This trait ensures that population inversion can be nearly obtained even when the pulse sequence has a short jump delay.
As an example, we employ the designed composite pulse sequence to prepare the $W$ state in a robust manner in the superconducting circuits.
The numerical results show that the fidelity can still maintain a high level in a decoherence environment.
\end{abstract}
\maketitle

\section{Introduction}
Robust control of quantum states is one of the crucial topics in quantum information processing (QIP)
\cite{PhysRevB.57.9024,PhysRevA.62.053409,PhysRevA.67.053404,PhysRevA.75.012329,
PhysRevA.102.013113,PhysRevA.102.022218,PhysRevA.102.042612,PhysRevApplied.14.064009}.
Through controlling the external fields, the initial state of a quantum system can be driven into the desired target state.
During this process, many requirements need to be satisfied.
The first one is to quickly accomplish quantum operations, because the coherence time of quantum systems is
generally very short.
Once the operation time is sufficiently long, the advantages of quantum computation would gradually be lost.
The second one is to perform quantum operations in a robust way.
As is well known, the infidelity of quantum logic gates needs to be below the fault-tolerant threshold ($10^{-4}$) to
protect the reliability of quantum computation \cite{Nielsen00}.
However, quantum systems always suffer from the harmful effect of external noises, which may be caused by
the imperfect knowledge of quantum systems or the fluctuations of external fields, etc.
{These external noises finally lead to a sharp decline in the fidelity of quantum gates.
In response, quantum control theory is developed to tackle these problems.}

Quantum control theory has been successfully applied in single
\cite{PhysRevB.89.245311,PhysRevLett.111.053603} and multibody systems
\cite{PhysRevLett.106.190501,PhysRevA.84.022326}.
At the same time,
this theory has made new progress in
quantum state preparations
\cite{PhysRevLett.116.230503,PhysRevLett.114.170501,PhysRevA.97.062343,PhysRevA.95.062319,PhysRevA.95.042318,PhysRevA.97.042122,Zhang2021},
quantum gate operations
\cite{PhysRevA.79.022301,PhysRevA.77.052303,PhysRevA.97.042336,PhysRevA.88.052326,PhysRevLett.124.220501,PhysRevLett.125.170502,PhysRevA.102.042607,PhysRevLett.125.250403,PhysRevA.101.032322,PhysRevApplied.14.034038,PhysRevA.101.052302,PhysRevA.94.022331},
and quantum error corrections
\cite{PhysRevA.101.052317,PhysRevA.101.023813,PhysRevA.102.013115,PhysRevLett.124.117701,PhysRevLett.124.153203,PhysRevLett.124.170501}.
{During the quantum control process, the primary mission is to design an appropriate pulse shape of the control fields to drive the
system evolution as we expect.
The traditional method is to employ the adiabatic passage (AP) \cite{1987Optical,PhysRevLett.68.2000,Melinger1994,PhysRevA.59.4494,PhysRevA.50.584,Vitanov2001,Goswami2003},
which matches the initial (target) state to one instantaneous eigenstate as an evolution path of the system.
The AP requests that the physical parameters must vary slowly enough so as to meet the adiabatic condition.
It is widely known that there are two relevant well-established techniques: the chirped rapid adiabatic passage (CHIRAP) \cite{PhysRevLett.68.2000,Melinger1994,PhysRevA.59.4494,PhysRevA.50.584} and the stimulated Raman adiabatic passage (STIRAP) \cite{Gaubatz1990,PhysRevA.44.R4118,PhysRevA.45.5297}.
The former proposes to apply a chirped pulse to modulate the frequency detuning of the system.
The slow chirp rate ensures that the adiabatic condition is strictly satisfied. Therefore, one can drive the system evolution along a given eigenstate and suppress transitions to other eigenstates \cite{PhysRevA.59.4494}.
The latter is one of the most popular methods for quantum control in three-level systems and has since been widely expanded to other fields \cite{RevModPhys.89.015006}.
In the STIRAP,
the Stokes pulse and the pump pulse are applied in a counterintuitive order,
and one can achieve a successful population transfer between two states under the two-photon resonance condition in three-level systems \cite{RevModPhys.89.015006}.
The main feature of AP is that the robustness against parameter fluctuations could be achieved at the expense of operation speed and accuracy.
However, once the adiabatic condition cannot be satisfied well,
the accuracy would be sharply dropped due to the imperfect adiabatic path.}

{To optimize the speed and accuracy of AP,
an improved and acclaimed technique named shortcut to adiabaticity (STA) \cite{RevModPhys.91.045001,PhysRevA.83.062116,PhysRevLett.105.123003} has been developed.}
{There are two common methods in the STA technique: Lewis-Riesenfeld invariants \cite{Lewis1969} and transitionless quantum driving \cite{Demirplak2003,Demirplak2005,Berry2009}.}
The main idea of STA is to design the pulse shape to canceling nonadiabatic transitions through an additional Hamiltonian.
Nevertheless, it depends heavily on the precisely known physical parameters, and strong fluctuations of parameters
might make this method inefficient.
{Recently, optimal control (OP) has become another popular way to achieve high-fidelity quantum operations} \cite{PhysRevLett.106.190501,PhysRevA.101.022320,PhysRevA.101.062307,PhysRevA.102.052605,PhysRevA.101.023410,PhysRevA.102.043707,PhysRevA.103.012404,Plantenberg2007,Glaser2015}.
One disadvantage of OP is that the pulse shape is always intricate and nonanalytic because it is
achieved by using advanced mathematical algorithms.
As a result, the design process only yields certain numerical solutions.
In order to take account of the robustness and accuracy of quantum control, one can turn to the composite
pulses (CPs) technique, where the physical parameters are flexible to design
and the shapes are readily implemented in experiments due to its commonality.

The CPs technique was born in the field of nuclear magnetic resonance
\cite{Wimperis1994,Levitt1986,Wimperis1991,Husain2013}.
Over the past decade,
CPs have provided a lot of solutions in quantum computation
\cite{PhysRevLett.113.043001,Ivanov2011,PhysRevA.99.013402,PhysRevA.83.053420,PhysRevA.87.052317,PhysRevA.101.013827,PhysRevA.101.012321} due to their extremely high accuracy,
robustness against errors, and extraordinary flexibility \cite{PhysRevA.103.033110}.
Generally speaking,
the CPs technique is composed of a series of precise constant pulses with different relative phases.
These pulses are usually imposed on external laser fields, electric fields, magnetic fields, or radio-frequency
fields, etc.
It is shown that the phase-modulated composite pulses (PMCPs) can not only improve the robustness against
errors (broadband pulse),
but also enhance the sensitivity and selectivity of excitation (narrowband pulse)
\cite{PhysRevA.102.013105}.
Recently, the PMCPs have also been designed to produce a rotation in the Bloch sphere
\cite{PhysRevA.99.013402}, where
the errors resulting from the pulse area are gradually compensated with the increase of the pulse number.
{Another application of PMCPs is to implement the NOT gate, and this scheme is robust against both offset uncertainties and
control field variations by a very small number of modulation parameters \cite{PhysRevA.101.012321}.}

However, when the system phase cannot be modulated or the system does not carry phase information, PMCPs would be invalid.
In order to supplement the deficiency of composite pulses,
researchers turn their attention to some other adjustable parameters.
Therefore,
the detuning-modulated composite pulses (DMCPs) and the strength-modulated composite pulses (SMCPs)
have become two additional promising methods.
Recently,
based on photonic integrated circuits,
Greener \emph{et al.} applied DMCPs in the coupled optical waveguide model to achieve complete light
transfer \cite{PhysRevA.100.032333}.
SMCPs are used to implement dynamically corrected single-qubit gates on singlet-triplet qubits
\cite{Wang2012,PhysRevA.89.022310}, where qubit manipulations are handled by adjusting the electrically controlled exchange coupling strength.
Note that most works \cite{PhysRevLett.113.043001,PhysRevA.99.013402,Ivanov2011,PhysRevA.83.053420,
PhysRevA.87.052317,PhysRevA.101.013827,PhysRevA.101.012321,
PhysRevA.103.033110,PhysRevA.102.013105,PhysRevA.89.022310,PhysRevA.100.032333,Wang2012} of CPs are based on two-level systems.
Recently, the CPs have been extended to the three-level systems and have provided many creative works
\cite{PhysRevResearch.2.043194,PhysRevA.103.052612,PhysRevResearch.2.043235,MansourzadehAshkani2021,PhysRevB.95.241307,Friesen2017,Shi2022},
such as the implementation of high-fidelity composite quantum gates \cite{PhysRevResearch.2.043194,PhysRevA.103.052612} and
the efficient detection of chiral molecules \cite{PhysRevResearch.2.043235}.
It is worth mentioning that those works \cite{PhysRevResearch.2.043194,PhysRevA.103.052612,PhysRevResearch.2.043235} still
focus on the phase modulation in three-level systems.

In this paper,
based on the composite pulses technique,
we develop a general SMCPs scheme for robust population inversion in
three-level systems.
According to the Taylor expansion, the final transition probability expression is rearranged into a series of error
terms.
With predetermined phases and pulse area,
these error terms could be effectively eliminated from low order to high order by properly adjusting coupling
strengths.
In addition,
we investigate the influence of waveform deformation on the transition probability.
{The results show that slight deformation still allows the SMCPs to work well, except for
the heavily deformed case.}
In the end,
we further apply the current SMCPs scheme to prepare the $W$ state in the superconducting quantum interference device (SQUID) model,
and the numerical simulations demonstrate that the SMCPs can still maintain a robust performance in a
decoherence environment.

The structure of this paper is organized as follows.
In Sec.~\ref{model},
we present the design procedure of the SMCPs in the three-level system,
which are directly derived from the total composite propagator.
In Sec.~\ref{cps},
we illustrate how to eliminate the effect of pulse area error by SMCPs with the specific number of pulses.
Here, the sequence of up to seven pulses has been studied,
and the longer sequence could also be obtained in a similar way.
Then,
we analyze the impact of the waveform deformation of CPs on the performance of the transition probability and { compare our scheme with other composite pulses}.
In Sec.~\ref{app},
we give the application of the SMCPs scheme in robustly preparing the $W$ state in the superconducting circuits.
Finally, we give the conclusion in Sec.~\ref{con}.

\section{Theoretical model}\label{model}

\begin{figure}[b]
\centering
\scalebox{0.75}{\includegraphics{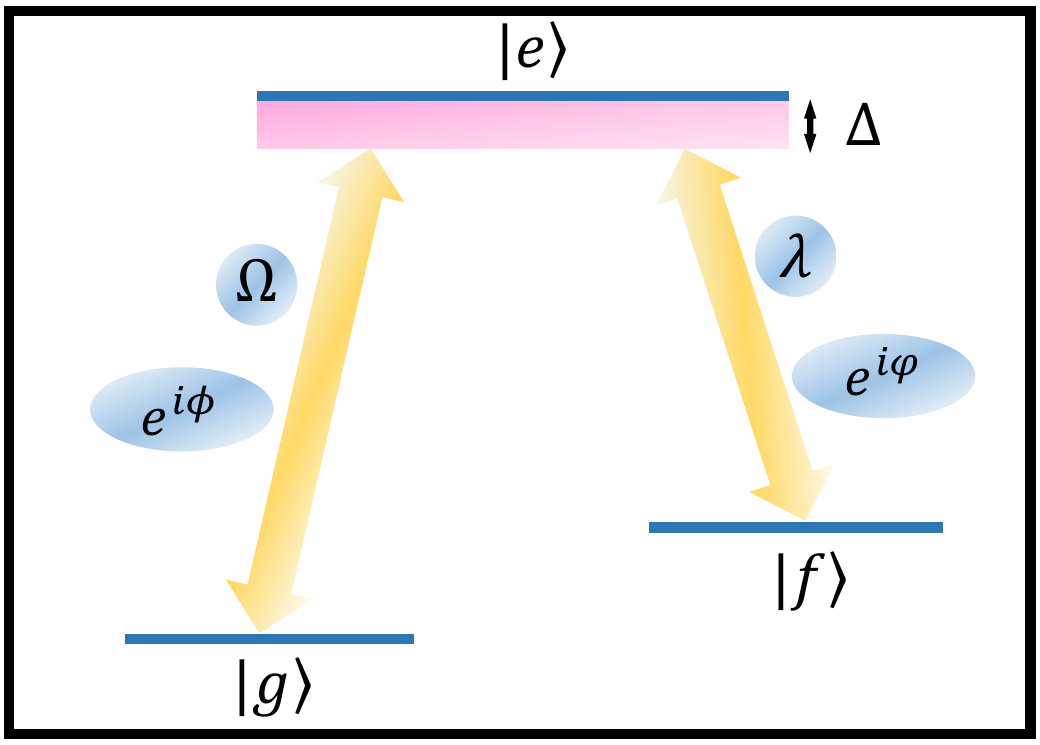}}
\caption{{A $\Lambda$-type three-level system driven by two control fields in the two-photon resonance regime.}}\label{mod}
\end{figure}

{In this section,
we elaborate on the design procedure of the general pulse waveform in a three-level system with
$\Lambda$-type structure.
This system has been regarded as a paradigmatic model in many branches of physics,
including atomic and molecular  \cite{PhysRevA.48.845,PhysRevA.53.373,PhysRevLett.87.183002},
quantum information science \cite{PhysRevA.73.042321},
and some other fields \cite{PhysRevLett.99.113003,PhysRevA.78.033416,Longhi2009,Wang2012(2)}.
The basic structure is shown in Fig.~\ref{mod},
where $|g\rangle$ and $|f\rangle$ are two ground states and $|e\rangle$ is the excited state.
There are two control fields that drive two transitions: $|g\rangle\leftrightarrow|e\rangle$ and $|f\rangle\leftrightarrow|e\rangle$.
Meanwhile, the transition frequency of the three-level system and the carrier frequency of the control fields satisfy the two-photon resonance condition \cite{scully97}. Note that the direct transition between two ground states is forbidden.
In the interaction picture,
the dynamics of the system can be described by the following Hamiltonian ($\hbar=1$ hereafter):
\begin{eqnarray}  \label{1}
H=\Delta|e\rangle\langle e|+\frac{1}{2} \Big(\Omega e^{i\phi}|g\rangle\langle e|+\lambda e^{i\varphi}|f\rangle\langle e|+\mathrm{H.c.}\Big),
\end{eqnarray}
where $\Omega$ and $\phi$ ($\lambda$ and $\varphi$) are the coupling strength and the phase of the transition
$|g\rangle\leftrightarrow|e\rangle $ ($|f\rangle\leftrightarrow|e\rangle$), and
$\Delta$ is the single-photon detuning. When the detuning is much larger than two coupling strengths (i.e., $|\Delta|\gg\Omega,\lambda$),
the excited state $|e\rangle$ could be adiabatically eliminated \cite{Brion_2007}.
Then, the three-level system would be reduced as an effective two-level system. In the following, we study the system dynamics including the excited state, and thus do not focus on this situation.

The propagator $U$ of this three-level system satisfies the following $\mathrm{Schr\ddot{o}dinger}$ equation:
\begin{eqnarray}
i \dot{U}=H U.
\end{eqnarray}
When the Hamiltonian given by Eq.~(\ref{1}) is time independent,
the solution of this equation becomes
$U=\mathrm{exp}\big({-i H T}\big)$,
where $T$ is the time duration.
It is instructive to adopt its matrix form in the basis $\{|g\rangle, |f\rangle, |e\rangle\}$,
which reads:
\begin{eqnarray}\label{ures}
U(\Theta)=
\left[
\begin{array}{ccc}	
U_{11}&U_{12} &U_{13} \\
U_{21}&U_{22} &U_{23}  \\
U_{31}&U_{32} &U_{33} \\
\end{array}	
\right],
\end{eqnarray}
where
\begin{eqnarray}
U_{11}&=&\cos^2{\Theta}+e^{-i\delta}(\cos{\frac{A}{2}}+\frac{i \Delta}{\sqrt{1+\Delta^2}}\sin{\frac{A}{2}})\sin^2{\Theta},\nonumber\\[0.2ex]
U_{22}&=&\sin^2{\Theta}+e^{-i\delta}(\cos{\frac{A}{2}}+\frac{i \Delta}{\sqrt{1+\Delta^2}}\sin{\frac{A}{2}})\cos^2{\Theta},\nonumber\\[0.2ex]
U_{33}&=&e^{-i\delta}(\cos{\frac{A}{2}}-\frac{i \Delta}{\sqrt{1+\Delta^2}}\sin{\frac{A}{2}}),\nonumber\\[0.2ex]
U_{12}&=&\frac{1}{2}e^{-i\delta}e^{i \Psi}(\cos{\frac{A}{2}}+\frac{i \Delta}{\sqrt{1+\Delta^2}}\sin{\frac{A}{2}}-e^{i\delta})\sin{2\Theta},
\nonumber\\[0.2ex]
U_{21}&=&\frac{1}{2}e^{-i\delta}e^{-i \Psi}(\cos{\frac{A}{2}}+\frac{i \Delta}{\sqrt{1+\Delta^2}}\sin{\frac{A}{2}}-e^{i\delta})\sin{2\Theta},
\nonumber\\[0.2ex]
U_{13}&=&-e^{-i 2 \delta}U_{31}^*=-i e^{-i\delta} e^{i \phi}\frac{\sin{\Theta}}{\sqrt{1+\Delta^2}}\sin{\frac{A}{2}},\nonumber\\[0.2ex]
U_{23}&=&-e^{-i 2 \delta}U_{32}^*=-i e^{-i\delta} e^{i \varphi}\frac{\cos{\Theta}}{\sqrt{1+\Delta^2}}\sin{\frac{A}{2}}. \nonumber
\end{eqnarray}
Here, $\Theta=\arctan{(\Omega/\lambda)}$, $\delta=\int _0 ^T\Delta/2 \  dt$, the phase difference $\Psi=\phi-\varphi$, and the expression of
the total pulse area $A$ is
\begin{eqnarray}\label{pluarea}
A=\int_0^T \sqrt{\Omega^2+\lambda^2+\Delta^2}\  dt.
\end{eqnarray}
Obviously, when the pulse area $A=2\pi$,
two ground states are decoupled from the excited state $|e\rangle$ so that we can only concentrate on the subspace $\{|g\rangle,|f\rangle\}$.
Then, when the system is initially in the ground state $|g\rangle$, the transition probability $P$ of the target state $|f\rangle$ becomes
\begin{eqnarray}\label{pf}
P =|U_{21}|^2=\cos^2{\left(\frac{\pi\Delta}{2\sqrt{1+\Delta^2}}\right)} \sin^2{2 \Theta}.
\end{eqnarray}
It is easily found from Eq.~(\ref{pf}) that
the detuning $\Delta$ is required to be zero in order to achieve complete population inversion for the single pulse.}

Then, one can see from Eq.~(\ref{pluarea}) that two kinds of uncertainties would cause the pulse area error when $\Delta=0$.
The first one is the coupling strength error,
which may be caused by the inhomogeneity of the control fields.
Another one is the inaccurate pulse duration,
which may be limited to the experiment condition or manual operation accuracy.
Note that tiny errors in the pulse area would have a significant effect on population inversion.
This can be demonstrated as follows.
Suppose that the pulse area error is represented by $\epsilon$,
and a deviation in the original pulse area reads $A=A(1+\epsilon)$.
As a result,
the transition probability given by Eq.~(\ref{pf}) is rewritten as
\begin{equation} \label{p5}
P(\epsilon)=\sin^4{\frac{A(1+\epsilon)}{4}} \sin^2{2 \Theta}.
\end{equation}
In order to study how the pulse area error makes an impact on population inversion,
the transition probability $P(\epsilon)$ given by Eq.~(\ref{p5}) is expanded as the Taylor series,
\begin{equation}\label{5}
\begin{split}
&P (\epsilon)=\alpha_{0}+\alpha_{1}\epsilon+\alpha_{2}\epsilon^2+O(\epsilon^3), \\
\end{split}
\end{equation}
where  $\alpha_{j}$ is the $j\mathrm{th}$-order coefficient, $j=0,1,2,\cdots$. Actually, $\alpha_{0}$ is the transition probability in the absence of the pulse area error.
By setting the pulse area $A=2\pi(1+\epsilon)$, the first three order coefficients are
 \begin{equation}\label{para}
\begin{split}
\alpha_{0}=\sin^2{2 \Theta},~\alpha_{1}&=0,~\alpha_{2}=-\frac{1}{2} \pi^2
\sin^2{2\Theta}.\\
\end{split}
\end{equation}
Notice that the first-order coefficient $\alpha_{1}$ is automatically eliminated because all odd-order coefficients contain the term $\sin{(A/2)}$.
Here,
$\Theta=\pi/4+k\pi/2$ is the indispensable condition to achieve population inversion, i.e.,
$\alpha_{0}=1$.
We can observe from Eq.~(\ref{para}) that the second-order coefficient $\alpha_{2}$ has the same monotony with $\alpha_{0}$ because they have the same term $\sin^2{2\Theta}$.
This means that the error would also reach maximum if we achieve complete population inversion.
\begin{figure}
\centering
\subfigure{\scalebox{0.41}{\includegraphics{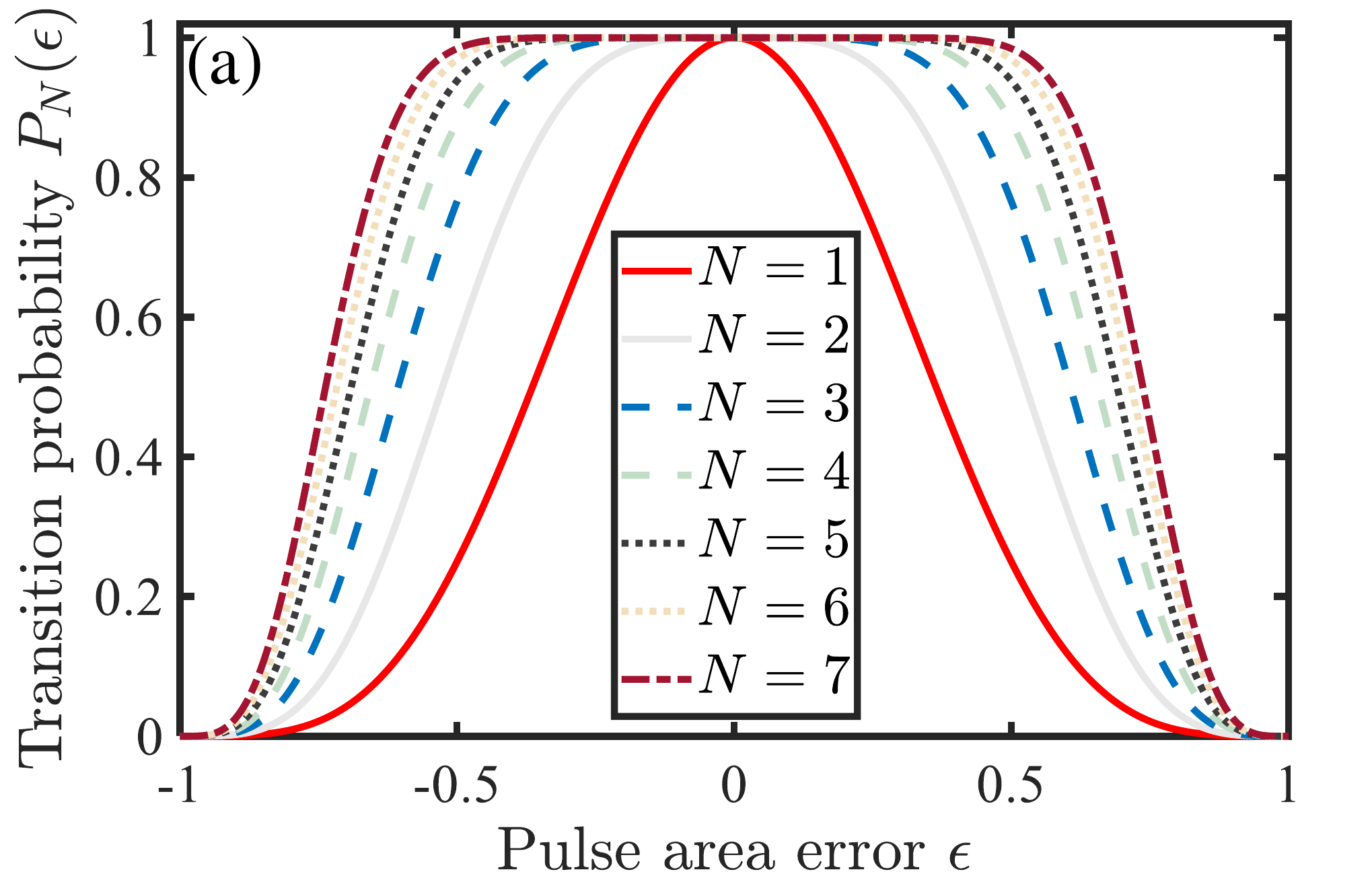}} \label{eps}}
\subfigure{\scalebox{0.41}{\includegraphics{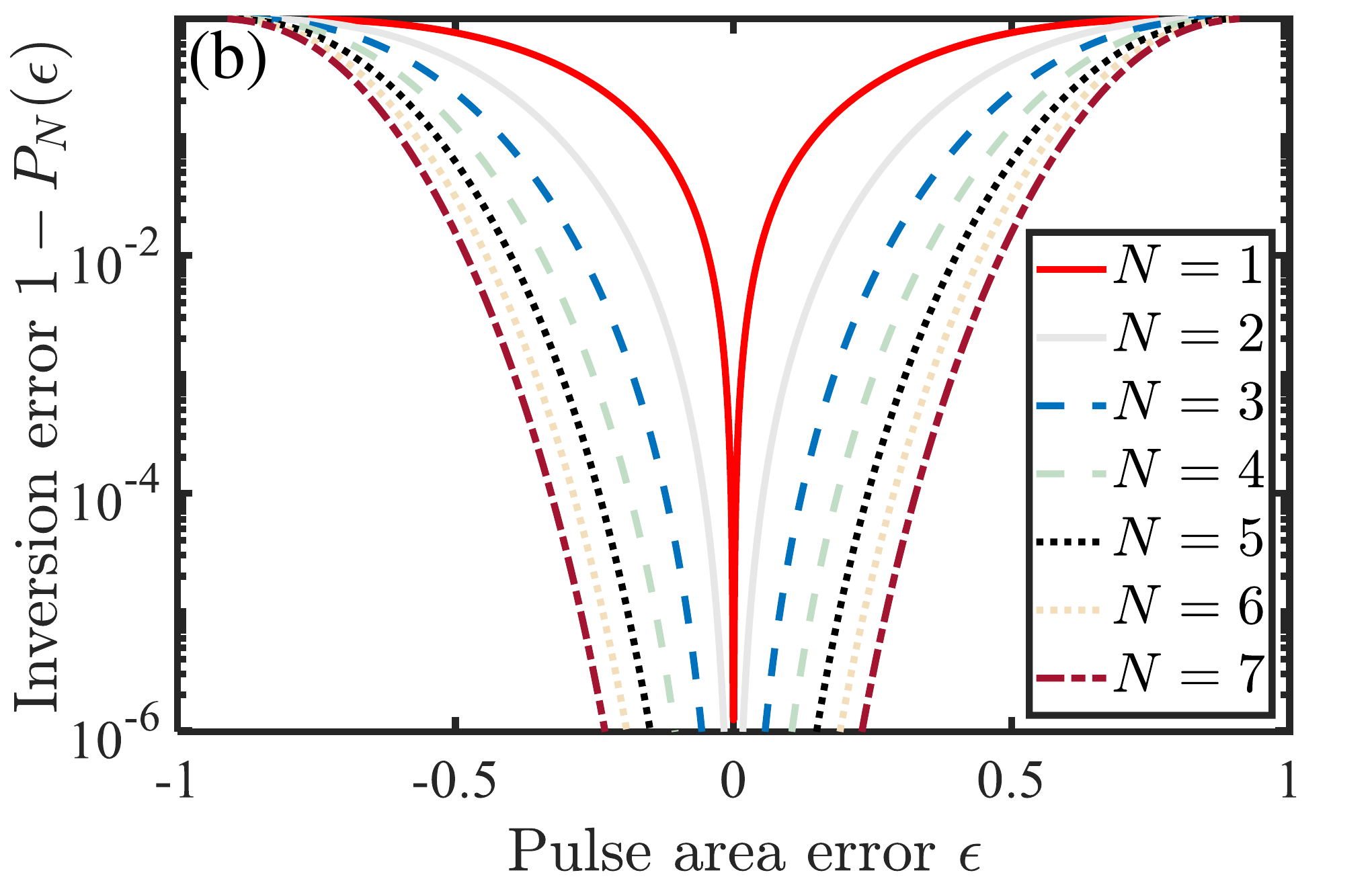}} \label{eps1}}
\caption{(a) Transition probability $P_N(\epsilon)$ vs the pulse area error $\epsilon$ under different pulses
sequences.
(b) Inversion error $1-P_N(\epsilon)$ vs the pulse area error $\epsilon$.
All $\Theta_n$ come from Table~\ref{tab1}.}
\end{figure}
In Figs.~\ref{eps} and \ref{eps1}, the red solid curve represents $P (\epsilon)$ as a function of the pulse area error $\epsilon$, which shows that  the transition probability sharply drops near $\epsilon=0$.
Hence, the single-pulse scheme does not possess a robust manner against the pulse area error.

{In order to solve this problem,
next we put forward the SMCPs scheme.
The sequence is composed of multiple single pulses,
where only coupling strengths for each pulse are different.
For the compact expression,
we label the Hamiltonian and the propagator of the $n\mathrm{th}$ pulse as
$H_n$ and $U_n(\Theta_n)$, respectively.
Then, the general form of the total propagator for the $N$-pulses sequence
can be written as
\begin{eqnarray}\label{UUU}
U^{(N)}\!=\!U_N(\Theta_N)U_{N-1}(\Theta_{N-1})\cdots U_2(\Theta_2) U_1(\Theta_1).
\end{eqnarray}
The detailed derivation of the total propagator $U^{(N)}$ without the pulse area error is presented in Appendix~\ref{A1}.
Note that the transition probability of every state is associated with both the detuning and the coupling strengths of each pulse.
For simplicity,
we assume each pulse is in the resonant regime ($\Delta_n = 0$) in the following.}

The target of this work is to achieve robust population inversion by SMCPs. Without loss of generality, we assume the initial state is $|\psi_i\rangle=|g\rangle$; then the target state becomes $|\psi_f\rangle=|f\rangle$.
Each pulse area is represented as $2\pi(1+\epsilon)$, and we label $P_N(\epsilon)$ as the transition probability of the state $|f\rangle$ in the $N$ pulses sequence.
Similar to the derivations in Eq.~(\ref{5}),
$P_N(\epsilon)$ can be written in the following form by the Taylor expansion:
\begin{equation}\label{taylor}
P_N(\epsilon)=\alpha_{N,0}+\alpha_{N,1}\epsilon+\alpha_{N,2}\epsilon^2+\cdots+O(\epsilon^n),
\end{equation}
where $\alpha_{N,j}$ is the $j\mathrm{th}$-order coefficient in the $N$ pulses sequence, and its expression can be deduced from the total propagator given by Eq.~(\ref{UUU}).

Here, we do not intend to change the pulse area and phases of this system,
and thus the pulse area and phases are predetermined by constants during the evolution process.
Note that all odd-order coefficients $\alpha_{N,j} (j=2n-1)$ would vanish if the pulse area of each pulse is equal to $2\pi$, i.e., $A_1=\cdots=A_N=2\pi$.
As a result, only even-order coefficients are left in Eq.~(\ref{taylor}).
By modulating the parameters $\Theta_n$ ($n=1,\dots,N$),
we demand that the zeroth-order coefficient be unity,
{which ensures complete population inversion,}
and other even-order coefficients vanish as many as possible.
{
Note that according to the expression $\Theta_{n}=\arctan{(\Omega_{n}/\lambda_{n})}$,
we can keep the coupling strength $\lambda_n$ unchanged and alter the coupling strength $\Omega_n$ to implement the modulation of $\Theta_n$.
This means that we only adjust one of the coupling strengths to achieve robust population inversion in the three-level system.}

To be specific, if the sequence is composed of $N$ pulses,
we first solve $N$ equations to obtain the solutions of $\Theta_n$,
i.e.,
\begin{numcases}{}\label{eqs}
\alpha_{N,0}=1,\nonumber\\
\alpha_{N,2}=0,\\
~~~\cdots \nonumber\\
\alpha_{N,2N-2}=0.\nonumber
\end{numcases}
{Then, we calculate different coupling strengths $\Omega_n$ of each pulse according to the expression $\Omega_n=\lambda_n\tan\Theta_n$.
Consequently, we have the composite pulses sequence to achieve population inversion. In this situation,
the transition probability is accurate to the order of $O(\epsilon^{2N-2})$.}
Compared with the single-pulse scheme,
the transition probability in the SMCPs scheme has a better robust manner against the pulse area error $\epsilon$ since the high-order coefficients disappear.
{In other words,
eliminating more high-order coefficients in Eq.~(\ref{taylor}) could further suppress the detrimental effect of the pulse area error.}
When the composite pulses sequence is sufficiently short, the analytical results are readily obtained by solving the corresponding equations.
However, the long pulse sequence would increase the number of equations. As a result, it is difficult to derive the analytical results.
Instead, we provide the numerical solutions for the long pulse sequence.
In order to demonstrate this issue in more detail,
we exemplify the SMCPs scheme with a specific number of pulses below.

\section{Strength-modulated composite pulses scheme}\label{cps}

In this section, we study how to design the SMCPs to achieve robust population inversion when the
pulse area and phases remain unchanged.
{For simplicity,
we uniformly set $A_n=2\pi$, $\phi_n=\pi/2$, $\varphi_n=0$, and assume the system always works in the resonance regime (i.e., $\Delta_n=0$)}.
Note that $\Theta_n$ are only the adjustable parameters.

\subsection{Two pulses}\label{cps2}

The propagator in the two-pulses sequence can be expressed by
\begin{eqnarray}  \label{10a}
U^{(2)}=U_2(\Theta_2)U_1(\Theta_1).
\end{eqnarray}
According to the propagator $U^{(2)}$ given by Eq.~(\ref{10a}), the transition probability $P_{2}(\epsilon)$ of the state $|f\rangle$ is
\begin{eqnarray}
P_{2}(\epsilon)&=&\frac{1}{4}\Big\{\sin^2{\!\pi\epsilon}\big[\cos{\!\Theta_1}\!\sin{\!\Theta_2}\! +\!\cos{\!\Theta_2}(2\sin{\!\Theta_1}\!+\!\sin{\!\Theta_2})\big] \cr\cr
&& \!-\!2\cos^4{(\frac{\pi\epsilon}{2})}\sin{2(\Theta_1\!-\!\Theta_2)}\Big\}^2.
\end{eqnarray}
The detailed derivation process is given in Appendix~\ref{AAA}.
By the Taylor expansion, $P_{2}(\epsilon)$ can be regrouped as
\begin{equation}
P_2(\epsilon)=\alpha_{2,0}+\alpha_{2,2}\epsilon^2+O(\epsilon^4),
\end{equation}
where the expressions of the first two coefficients read
\begin{eqnarray}
\alpha_{2,0}&=&\sin^2\big[2(\Theta_1-\Theta_2 )\big],\cr\cr
\alpha_{2,2}&=&\pi ^2 \sin 2 (\Theta_2-\Theta_1) \Big[\sin 2 (\Theta_1-\Theta_2)\cr\cr
&&+2\cos \Theta_2 \sin \Theta_1+\frac{1}{2}\sin 2\Theta_1+\frac{1}{2}\sin \Theta_2\Big]. \nonumber
\end{eqnarray}

In the two-pulses sequence, we need to simultaneously solve the following equations:
\begin{numcases}{}
\alpha_{2,0}=1, \label{two1}\\
\alpha_{2,2}=0. \label{two2}
\end{numcases}
After some calculations, one solution of Eqs.~(\ref{two1}) and (\ref{two2}) can be written as
\begin{equation} \label{PHIE}
\begin{split}
\Theta_1=m\pi+\frac{\pi}{8}, ~\Theta_2=l\pi+\frac{3\pi}{8},
\end{split}
\end{equation}
where $m$ and $l$ are arbitrary integers.
By this choice of $\Theta_1$ and $\Theta_2$, the transition probability is accurate to the fourth order in the pulse area error $\epsilon$,
\begin{equation}
P_{2} (\epsilon)=1+O(\epsilon^4).
\end{equation}
The gray solid curve in Fig.~\ref{eps} represents the relation between the transition probability and the pulse area error when $\Theta_1=\pi/8$ and $\Theta_2=3\pi/8$. The result shows that the transition probability appears as a small flat top profile against the pulse area error in the two-pulses sequence.
Correspondingly, the gray solid curve in Fig.~\ref{eps1} demonstrates that the inversion error in the two-pulses sequence is much lower than that in the single pulse under the same condition.
Obviously,
the two-pulses sequence has a better fault tolerance than the single pulse in the error-prone environment.

{
Note that the above designed sequence can also used for implementing a NOT gate with a well-defined phase.
Although
the phases are inessential for population inversion, they play a very important role when achieving a quantum gate because different phases would determine different types of quantum gates.
For our objective, by the group of solution $\Theta_1=\pi/8$ and $\Theta_2=3\pi/8$ and the phase difference $\Psi_n=\pi/2$, the final propagator in the basis $\{|g\rangle, |f\rangle\}$ becomes (up to a global phase)
\begin{eqnarray}
U^{(2)}=
\left[
\begin{array}{ccc}	
0&1  \nonumber\\
1&0 \nonumber \\
\end{array}	
\right],
\end{eqnarray}
which is actually the $X$ gate. However, when the phase difference is chosen as $\Psi_n=\pi$, the final propagator would become
 \begin{eqnarray}
U^{(2)}=
\left[
\begin{array}{ccc}	
0&-i  \nonumber\\
i&0 \nonumber \\
\end{array}	
\right],
\end{eqnarray}
which is the $Y$ gate; see Appendix \ref{AAA} for details.

{
Furthermore,
this current sequence is readily extended to yield an arbitrary rotation gate.
The general form of an arbitrary rotation gate in the basis $\{|g\rangle, |f\rangle\}$ is \cite{Nielsen00}
\begin{eqnarray}
\hat{R}=\left[
\begin{array}{cc}	
\cos{(\theta/2)}e^{i \beta'}&\sin{(\theta/2)e^{i\beta}} \nonumber\\
-\sin{(\theta/2)}e^{-i\beta}&\cos{(\theta/2)}e^{-i \beta'}  \nonumber \\
\end{array}	
\right],
\end{eqnarray}
where $\theta$ is the rotation angle ($0\leq\theta\leq\pi$), and
$\beta$ and $\beta'$ are relative phases.
To implement this rotation gate,
Eq.~(\ref{two1}) needs to be modified as $\alpha_{2,0}=\sin^2{(\theta/2)}$, and then we properly adjust the phases of the coupling strengths.
As an example, to implement the Hadamard gate by the two-pulses sequence,
we first need to obtain a maximum superposition state of $|g\rangle$ and $|f\rangle$, and the equations read
\begin{numcases}{}
\alpha_{2,0}=1/2, \nonumber \\
\alpha_{2,2}=0. \nonumber
\end{numcases}
One group of the solutions is $\Theta_1=0.1047$ and $\Theta_2=0.4974$.
Then,
we need to choose two proper phases $\phi_n=\varphi_n$, and obtain the Hadamard gate
\begin{eqnarray}
H=
\frac{1}{\sqrt{2}}\left[
\begin{array}{cc}	
1&1 \nonumber\\
-1&1 \nonumber \\
\end{array}	
\right].
\end{eqnarray}
Note that different from the previous work~\cite{PhysRevResearch.2.043194} for the Hadamard gate in the three-level system,
the total pulse area of the current sequence can be $4\pi$, while it is $6\pi$ for the sequence $\mathcal{H}_6$ in Ref.~\cite{PhysRevResearch.2.043194}.
This flexibility of the sequence length benefits from the current sequence starting straight from the three-level system rather than nesting the existing sequence. }

}

\subsection{Three pulses}

The three-pulses sequence produce the following propagator:
\begin{eqnarray}
U^{(3)}=U_3(\Theta_3)U_2(\Theta_2)U_1(\Theta_1).
\end{eqnarray}
Thus, the exact expression of the transition probability $P_3(\epsilon)$ can be written as
\begin{widetext}
\begin{eqnarray}
P_3(\epsilon)&=&\Big\{\!\sin\!2\Theta_3(\sin\!2\Theta_1\sin\!2\Theta_2\!- \!2\sin\!\Theta_1\sin\!\Theta_2\sin\!\pi\epsilon\sin\!2\pi\epsilon)\!+ \!8\cos^2\!\Theta_1\!\big[\!\cos\!\Theta_2\sin\!\Theta_3\cos(\Theta_2\!-\!\Theta_3)\!+ \!\sin\!\Theta_2\cos\!\Theta_3\sin^2\!\pi\epsilon\big]\!\nonumber\\[0.1ex]
&&+4\sin^2\!\Theta_3\!\big[\sin\!2\Theta_1\sin^2\!\Theta_2\!+\!2\sin\!\Theta_1\cos\!\Theta_2\sin^2\!\pi\epsilon\big]\!- \!4\sin^2\!\pi\epsilon(\sin\!2\Theta_1\!\cos\!\Theta_2\!\cos\!\Theta_3\!- \!4\sin\!\Theta_1\sin\!\Theta_2\sin\!2\Theta_3)\!\nonumber\\[0.1ex]
&&-\cos\pi\epsilon\big[\!\sin2\Theta_1\!+\!\sin2(\Theta_1\!-\!\Theta_2)\!+ \!\sin2(\Theta_1\!-\!\Theta_3)\!-\!3\sin\!2(\Theta_1\!-\!\Theta_2\!+\!\Theta_3)\!+ \!4\sin\Theta_2\cos\Theta_3\cos(\Theta_2\!-\!\Theta_3)\nonumber\\
&&+8\sin\Theta_1\cos\Theta_3\sin^2\pi\epsilon(\cos(\Theta_1\!-\!\Theta_2)\!+ \!\cos\Theta_2\!\cos\Theta_3\!+\!1)\!\big]\!-\!\cos^2\!\pi\epsilon\big[\!\sin2\Theta_1\!+ \!\sin2\Theta_2\!+\!\sin2\Theta_3\!-\!\sin2(\Theta_1\!-\!\Theta_2)\!\nonumber\\[0.1ex]
&&-\!\sin2(\Theta_1\!-\!\Theta_3)\!-\!\sin2(\Theta_2\!-\!\Theta_3)\!- \!3\sin2(\Theta_1\!-\!\Theta_2\!+\!\Theta_3)\!\big]\!+\!8\cos^3\!\pi\epsilon\sin\!\Theta_1\cos \Theta_3\cos(\Theta_1\!-\!\Theta_2)\cos (\Theta_2\!-\!\Theta_3)\Big\}^2. \nonumber
\end{eqnarray}
Similarly, through the Taylor
expansion, the transition probability $P_{3}(\epsilon)$ can be expanded as
\begin{equation}
P_{3}
(\epsilon)=\alpha_{3,0}+\alpha_{3,2}\epsilon^2+\alpha_{3,4}\epsilon^4+O(\epsilon^6).
\end{equation}
Since only three parameters $\{\Theta_1, \Theta_2, \Theta_3\}$ are contained in the three-pulses sequence, it is sufficient to extend Eq.~(\ref{taylor}) to the fourth-order term.
The expressions of the first three coefficients are
\begin{eqnarray}
\alpha_{3,0}&\!=\!&\sin^2{2( \Theta_1-\Theta_2+\Theta_3)},\nonumber\\[0.1ex]
\alpha_{3,2}&\!=\!&-\frac{\pi^2}{2}\sin{2(\Theta_1\!-\!\Theta_2\!+\!\Theta_3)}\Big\{\sin{2(\Theta_1\!-\!\Theta_2)}\!+\!2\cos{(\Theta_1\!-\!\Theta_2)}\sin{(\Theta_1\!+\!\Theta_2\!-\!2\Theta_3)}\!+\!3\sin{2(\Theta_1\!-\!\Theta_2\!+\!\Theta_3)}\!\cr
&&+4\sin{\Theta_1}\cos{(\Theta_2\!-\!2\Theta_3)}\!+\!4\cos{\Theta_3}\big[\sin{(2\Theta_1\!-\!\Theta_2)}\!+\!\sin{\Theta_1}\big]\Big\}, \\
\alpha_{3,4}&\!=\!&\frac{\pi^4}{48}\Big\{\!48\big[\sin\!\Theta_1\cos(\Theta_2\!-\!2\Theta_3)\!+\!\sin(2\Theta_1\!-\!\Theta_2)\cos\Theta_3+\sin\!\Theta_1\cos\Theta_3\!+\!\cos(\Theta_1\!-\!\Theta_2)\sin(\Theta_1\!-\!\Theta_3)\cos(\Theta_2\!-\!\Theta_3)\cr
&&+51\sin^2\!2(\Theta_1\!-\!\Theta_2\!+\!\Theta_3)\big]^2\!\!+\!2\sin\!2(\Theta_1\!-\!\Theta_2\!+\!\Theta_3)\big[64\sin\!\Theta_1\cos(\Theta_2\!-\!2\Theta_3)\!+\!4\cos\!\Theta_3\big[16\sin(2\Theta_1\!-\!\Theta_2)\!+\!19\sin\!\Theta_1\cr
&&+3\sin\!\Theta_2\big]\!+\!64\cos(\Theta_1\!-\!\Theta_2)\sin(\Theta_1\!-\!\Theta_3)\cos(\Theta_2\!-\!\Theta_3)\!+\!12\sin\!\Theta_1\!\cos\!\Theta_2\!+\!3(\sin\!2\Theta_1\!+\!\sin\!2\Theta_2\!+\!\sin\!2\Theta_3)\big]\Big\}.\nonumber
\end{eqnarray}
\end{widetext}

{Therefore, in the three-pulses sequence,
the solutions need to simultaneously satisfy the following equations:}
\begin{numcases}{}
\alpha_{3,0}=1, \label{3peq}\\
\alpha_{3,2}=0,\label{3peq1}\\
\alpha_{3,4}=0. \label{3peq2}
\end{numcases}
Here, the solution for Eq.~(\ref{3peq}) can be written as
$\Theta_1\!-\!\Theta_2\!+\!\Theta_3\!=\!\pi/4+m\pi/2$, where $m$ is an arbitrary integer.
However, it is difficult to derive analytical formulas for Eqs.~(\ref{3peq1}) and (\ref{3peq2}). Instead, we employ the numerical method
to obtain the solutions of Eqs.~(\ref{3peq})--(\ref{3peq2}). One numerical solution is presented in Table~\ref{tab1}.
We figure out the profile of the transition probability $P_{3}(\epsilon)$ by the blue dashed curves in Figs.~\ref{eps} and \ref{eps1}.
{Clearly,
the three-pulses sequence produces a wider top platform than the two-pulses
sequence because the fourth-order coefficient $\alpha_{34}$ vanishes.}
In this situation, the transition probability is accurate to the sixth order in the pulse area error $\epsilon$,
\begin{equation}
P_{3} (\epsilon)=1+O(\epsilon^6).
\end{equation}

\begin{table}[htbp]
\caption{Values of the coupling strength ratio to achieve population inversion.\label{tab1}}
 \centering
 \begin{tabular}{c|c|c|c|c|c|c}
 \hline
  \hline
  \backslashbox{\kern+2em}{$N$\kern+1em}&2&3&4&5&6&7\cr
  \hline
$\Theta_1$                    &0.3927                    ~&0.1944                  ~&0.0897                    ~&0.0477
~&0.0203                   ~&0.0137\cr
  \hline
$\Theta_2$                    &1.1781                    &0.7854                  &0.4688                    &0.2913
&0.1508                   &0.1072\cr
  \hline
$\Theta_3$                    &--&1.3764                   &1.0613                   &0.7854                   &0.5004
&0.3811\cr
  \hline
$\Theta_4$                    &--&--&1.4676                   &1.2795                   &0.9948                   &0.8233\cr
  \hline
$\Theta_5$                    &--&--&--&1.5231                   &1.3826                   &1.2459\cr
  \hline
$\Theta_6$                    &--&--&--&--&1.5431                   &1.4854\cr
  \hline
$\Theta_7$                    &--&--&--&--&--&1.5606\cr
  \hline
  \hline
 \end{tabular}
\end{table}

\subsection{More than three pulses}\label{morethan}

Next, we briefly present the multiple-pulses design procedure.
The derivations of the sequence with larger pulse numbers are similar to the two-(three)-pulses sequence.
At first, the total propagator of the $N$-pulses sequence is
\begin{eqnarray}
U^{(N)}=U_N(\Theta_N)U_{N-1}(\Theta_{N-1})\dots U_1(\Theta_1).
\end{eqnarray}
In the $N$-pulses sequence,
the transition probability could be expanded to $(2N)\mathrm{th}$ order by the Taylor expansion
\begin{equation}
P_{N}(\epsilon)=\alpha_{N,0}+\alpha_{N,2}\epsilon^2+\alpha_{N,4}\epsilon^4+\dots+O(\epsilon^{2N}).
\end{equation}
It is not hard to find that the zeroth coefficient $\alpha_{N,0}$ can be expressed by (see Appendix \ref{A1} for details)
\begin{equation}  \label{28a}
\alpha_{N,0}=\sin^2{\bigg[2\sum^{N}_{k=1}(-1)^{k+1}\Theta_k\bigg]}.
\end{equation}
Therefore, the solution of the equation $\alpha_{N,0}=1$ can be written as
\begin{eqnarray}
\sum^{N}_{k=1}(-1)^{k+1}\Theta_k=\frac{\pi}{4}+\frac{m\pi}{2}, \nonumber
\end{eqnarray}
where $m$ is an arbitrary integer.

Then, the remaining
$(N-1)$ equations to be solved come from Eqs.~(\ref{eqs}).
In this situation,
the maximum transition probability is accurate to the $(2N)\mathrm{th}$ order in the pulse area error $\epsilon$,
\begin{equation}
P_{N} (\epsilon)=1+O(\epsilon^{2N}).
\end{equation}
We present some numerical solutions of $\Theta_n$ in Table \ref{tab1}, and plot in Figs.~\ref{eps} and \ref{eps1} the performance of the maximum transition probability produced by the four pulses to the seven pulses.
As expected, the more pulses we apply,
the broader region of the high fidelity against the pulse area error we will get.
For example,
the inversion error $1-P_{N}(\epsilon)$ still remains below $10^{-4}$ even though the pulse area error reaches 0.328 in the seven-pulses sequence.

\subsection{The influence of waveform deformation}\label{wfde}

In the ideal situation,
the waveform of CPs is the perfect square wave.
{Unfortunately,
the limitations of experimental conditions would produce imperfect pulse shapes in practice.}
The main problem is the waveform deformation,
which usually takes place at the moment of the square wave switching,
reflected in the strength jump delay,
i.e., the rise and fall edges.
As a result,
the quantum system is approximately driven by a time-dependent pulse instead of constant pulses sequence.
Here,
taking the three-pulses sequence for example,
we study this issue by designing a function to simulate the imperfect square pulse
waveform, and the expression of the coupling strength is
\begin{eqnarray}\label{27a}
&\Omega(t)= \left \{
\begin{array}{ll}
    \displaystyle\Omega_{2}\!-\!\frac{\Omega_{2}\!-\!\Omega_{1}}{1\!+\!e^{\tau(t\!-\!t_1)}},           &
    \displaystyle~~~ t \leq \frac{T\!+\!t_1\!-\!t_3}{2}, \\[2.8ex]
    \displaystyle\Omega_{3}\!-\!\frac{\Omega_{3}\!-\!\Omega_{2}}{1\!+\!e^{\tau(t\!-\!t_1\!-\!t_2)}},     &
    \displaystyle~~~ \frac{T\!+\!t_1\!-\!t_3}{2}<t \leq T,~~ \\
\end{array}
\right.
\end{eqnarray}
where $\tau$ is called the waveform deformation parameter and $t_n=2\pi/\sqrt{\Omega_n^2+\lambda_n^2}$, $n=1,2,3$.
\begin{figure}[htbp]
\centering
\subfigure{\scalebox{0.40}{\includegraphics{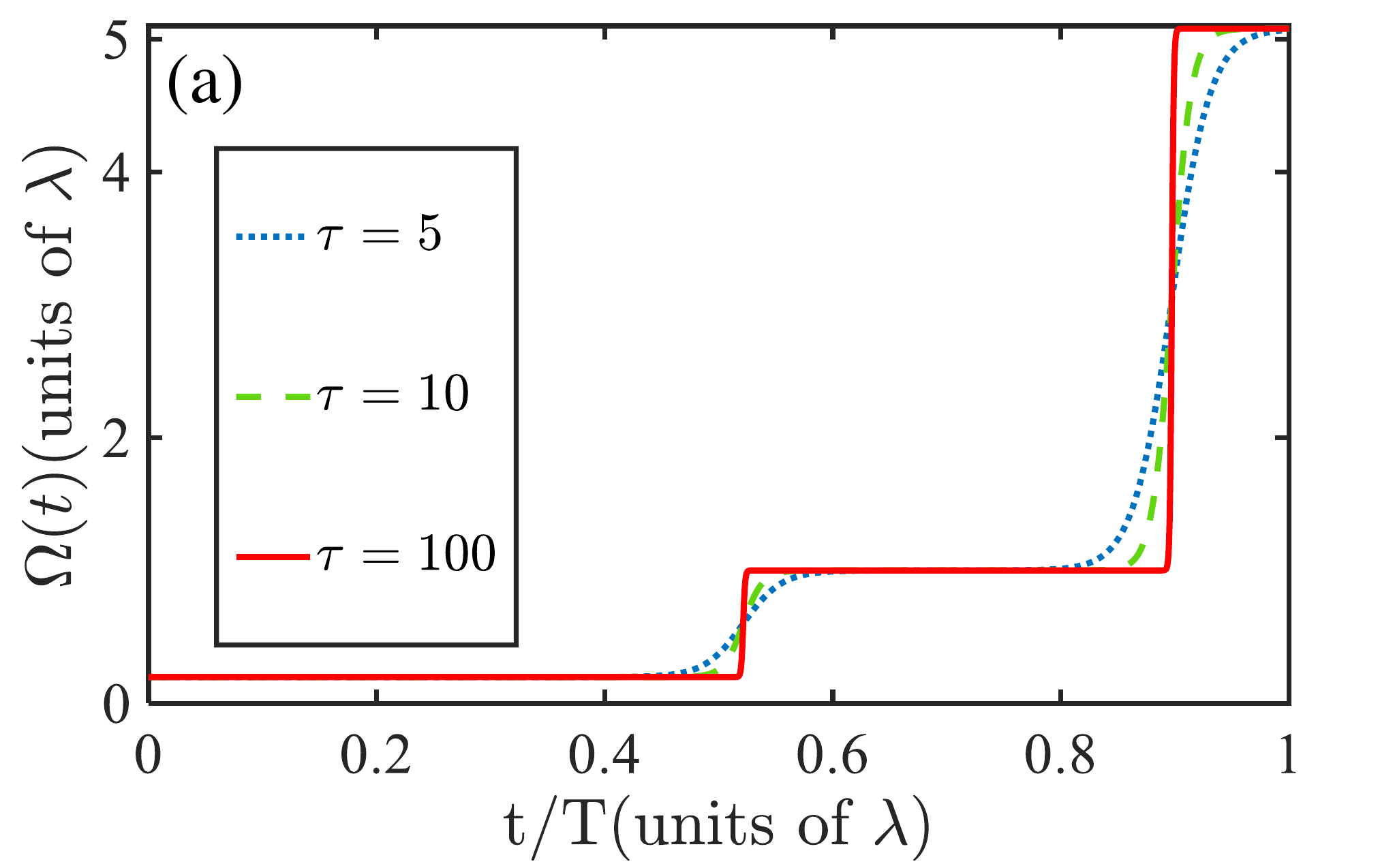} \label{Robst}} }
\subfigure{\scalebox{0.34}{\includegraphics{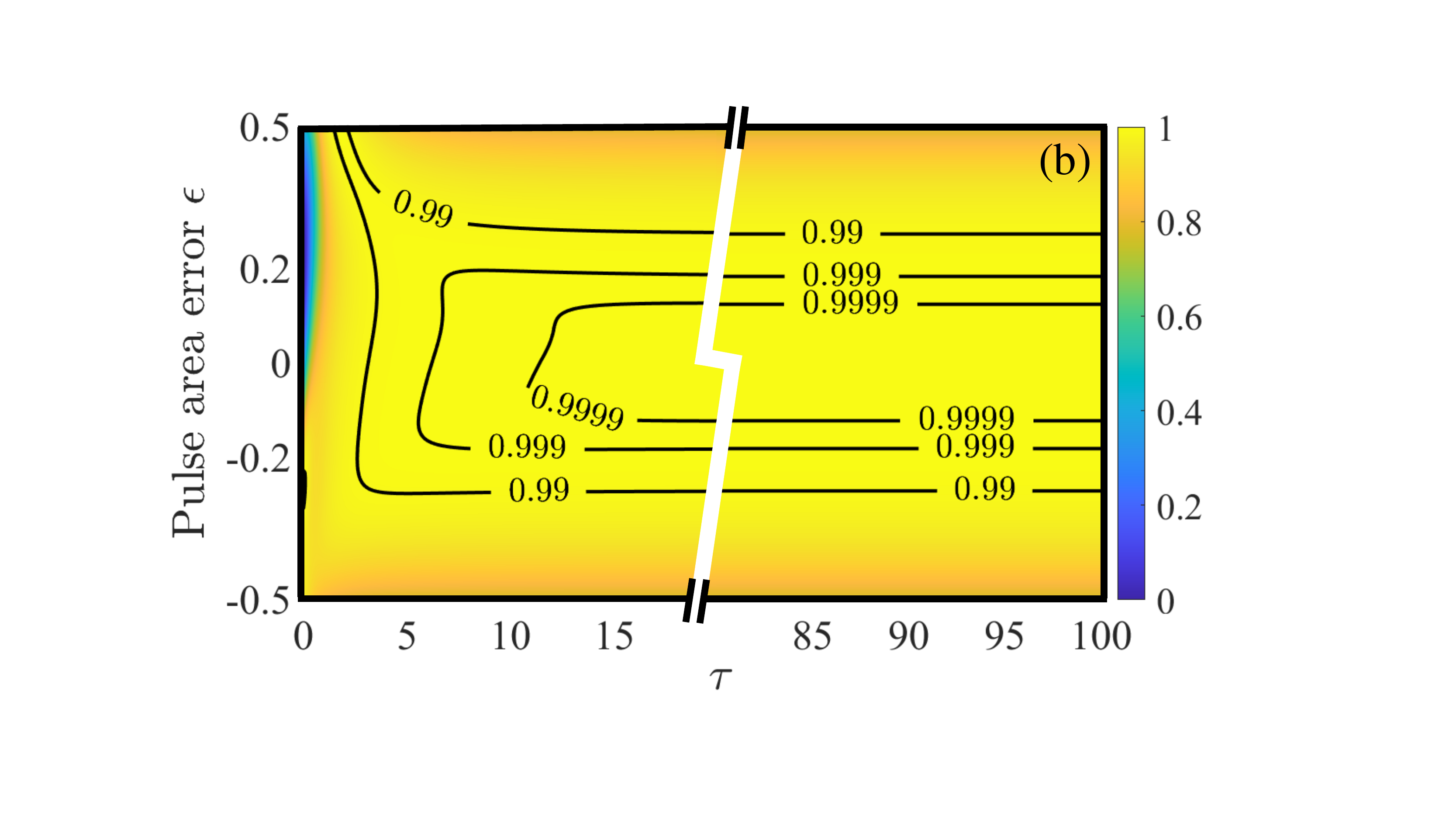} \label{Robst1}} }
\caption{(a) The waveform of the coupling strength $\Omega(t)$ with different deformation
degree in the three-pulses sequence.
(b) The maximum transition probability $P_3(\epsilon)$ vs the deformation parameter $\tau$ and the pulse area error
$\epsilon$. All $\Theta_n$ come from Table \ref{tab1}.
}
\end{figure}

It is easily found from Eq.~(\ref{27a}) that
one can control the width of the rising and falling edges by adjusting the dimensionless parameter $\tau$.
When the value of $\tau$ is larger,
the sequence given by Eq.~(\ref{27a}) is much closer to the constant pulses sequence.
For example,
it is shown by the red solid curve in Fig.~\ref{Robst} that
the deformation degree of the square wave is already negligible when $\tau=100$.
When $\tau \rightarrow +\infty$, Eq.~(\ref{27a}) describes the perfect square wave. However, when the value of $\tau$ is sufficiently small, the waveform described by Eq.~(\ref{27a}) is quite different from the perfect square wave, as shown by the blue dotted curve in Fig.~\ref{Robst}.
We plot in Fig.~\ref{Robst1} the transition probability as a function of the pulse area error and the waveform deformation.
The results demonstrate that a slight waveform deformation has little influence on our scheme since the width of the high transition probability region almost remains unchanged when $\tau> 10$.
A noticeable change appears in the transition probability only when the waveform deformation is very severe. However, the transition probability still maintains at a high level. Hence, our scheme is also robust against the waveform deformation.

{\subsection{Comparison with other composite pulses}
In this section,
we make a brief comparison of our CPs sequence with some previous works.
First,
most previous works \cite{Husain2013,PhysRevA.99.013402,PhysRevA.87.052317,PhysRevA.83.053420,PhysRevResearch.2.043194} are based on phase modulation,
in which the phase is a time-dependent variable parameter.
Different from the phase modulation,
here we propose a distinct modulation method: coupling strength modulation,
where all phases of the system remain unchanged during the evolution process.
Second,
the detuning modulation is equivalent to the strength modulation, in principle,
since two categories of the Hamiltonian can be associated through a rotation transformation \cite{SHEVCHENKO20101}.

{
The common starting point of the current sequence and the sequence in Ref. \cite{PhysRevResearch.2.043194} is to provide robustness against the pulse area error in the three-level system.
With the same total pulse area,
the robustness of our composite sequence agrees with that of the composite sequence in Ref.~\cite{PhysRevResearch.2.043194}. However, there are some differences between them.
At first, the structure of the sequences is distinct. We have only three pulses, while there are six pulses in Ref.~\cite{PhysRevResearch.2.043194} with the same pulse area.
From the perspective of the waveform,
our sequence is a simpler one because it contains fewer pulse jumps.
Furthermore, the design procedures are quite different. The sequence in Ref.~\cite{PhysRevResearch.2.043194} is constructed by the combination of a preexisting CPs sequence in two-level systems \cite{PhysRevA.83.053420}, ignoring the dynamics of the excited state.
Here, we consider the dynamics of all states in the three-level system, and directly design the composite pulses sequence according to the propagator.
Finally, the application systems are different, since we use strength modulation here, while the sequence in Ref.~\cite{PhysRevResearch.2.043194} relies on phase modulation.
To be specific,
one could adopt the phase modulation in some systems with easily adjustable phases (e.g., the trapped ions system \cite{Ivanov2011})
while the strength modulation could be conveniently applied in systems with uncontrollable phases (e.g., the integrated photonic circuits \cite{Politi2008}).}

For the strength modulation,
the difference between the two-level system and the three-level system is mainly embodied in the number of modulation parameters.
There are two adjustable coupling strengths in the three-level system,
while there is only single coupling strength in the two-level system.
When the pulse area is fixed,
one cannot eliminate the error by only modulating the coupling strength in a two-level system \cite{PhysRevA.99.013402}.
In order to make this strength modulation valid,
one must also modulate the pulse area,
as done in Ref.~\cite{PhysRevA.89.022310}.
However,
this situation is quite different in the three-level system.
Without a change in the pulse area,
we can modulate the ratio of two coupling strengths to effectively eliminate the systematic errors.

Moreover,
the SMCPs in Ref.~\cite{PhysRevA.89.022310} focuses on altering the exchange coupling to achieve robust quantum control in the singlet-triplet qubit.
This kind of pulse sequence needs to nest amounts of short pulses.
Hence, the pulse length would be long and may not maintain a high-fidelity performance in the decoherence environment.
In our scheme,
the pulse length can be appropriately chosen to equilibrate between pulse duration and robustness.
On the other hand,
the SMCPs in Ref.~\cite{PhysRevB.95.241307} are designed for suppressing leakage in the three-level system.
This scheme \cite{PhysRevB.95.241307} is required to simultaneously include two modulation parameters,
including the on-site potentials and the tunneling coupling.
In the current scheme,
only one of the coupling strengths is sufficient to be modulated to implement robust quantum control, while another coupling strength remains unchanged during the evolution process.}

\section{Illustrative example}\label{app}

{In this section,
we mainly illustrate the applications of the current SMCPs sequence for concrete physical models.
This current sequence is suitable for some systems where the phase is hardly to be modulated.
For example,
in integrated photonic circuits \cite{Politi2008}, since the coupling parameters between different waveguides are real valued,
the traditional phase modulation CPs is unapplicable.
During the light transportation process from one waveguide to another,
one can alter the spatial distance between waveguides, which is recognized as one kind of strength modulation between different waveguides.
And the amount of light coupling from one waveguide to the other can be tuned by lithographically adjusting their width \cite{Politi2008}.
Thus, one can adopt strength modulation CPs to achieve robust light transportation between different waveguides  in integrated photonic circuits  \cite{PhysRevA.100.032333}.

The current sequence is also feasible for the system where the strength parameters are conveniently modulated.
For instance,
in the system consisting of multiple quantum dots \cite{PhysRevLett.116.110402},
there are two strength parameters involved in the control of quantum dots at the same time.
One is the chemical potential of quantum dots, which is used for modulating the detuning.
The other is the tunneling barrier between adjacent quantum dots,
which alters the coupling of quantum dots.
Note that both strength parameters can be appropriately controlled by adjusting gate voltages, taking advantage of short-range interaction and electrical controllability \cite{PhysRevLett.116.110402}.
Therefore, a proper value of the gate voltage would make two quantum dots couple efficiently and produce a time-dependent exchange.
So it is convenient for employing the strength modulation in this system.

In the superconducting circuits system \cite{RevModPhys.85.623},
both the strength modulation and the phase modulation can be used to achieve robust quantum control by modulating the interaction of external fields.
When considering the weak dephasing in this system,
the phase modulation method \cite{PhysRevA.103.033110} is less robust against the phase noise than the strength modulation method.
Thus,
systematic errors may not be effectively eliminated due to the dephasing, and this problem can be elegantly circumvented by
the current SMCPs sequence.
In the following,
we choose this system to illustrate how to achieve robust quantum control by our SMCPs sequence.}

\begin{figure}[t]
\centering
\scalebox{0.31}{\includegraphics{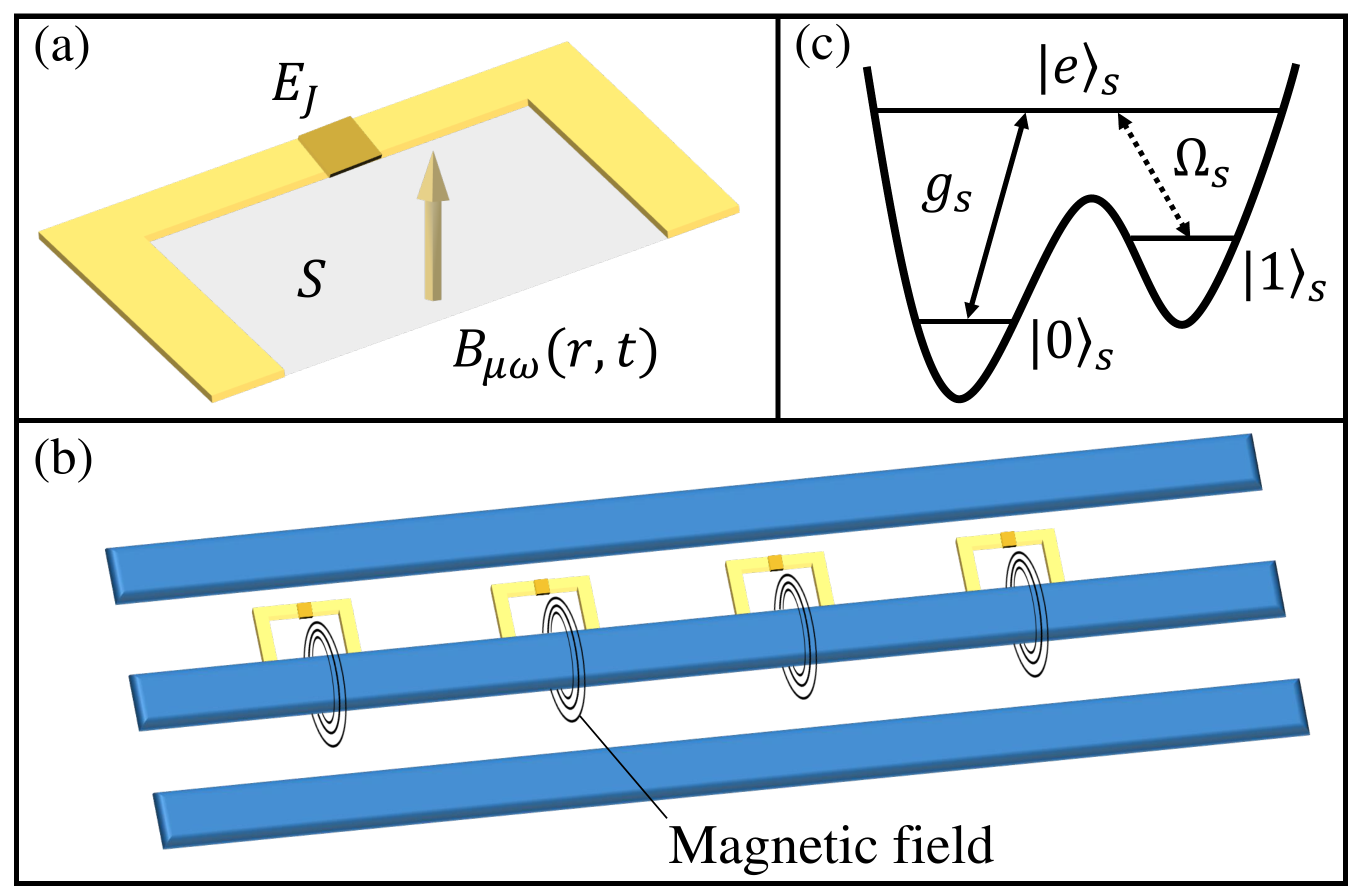}}
\caption{(a) Schematic diagram of the SQUID qubit. (b) Four SQUID qubits coupled to a single-mode cavity field.
 (c) The energy-level configuration of every SQUID qubit.
}\label{squid}
\end{figure}

The superconducting quantum interference devices (SQUIDs) are elementary components in superconducting circuits \cite{Xia2008,PhysRevA.67.042311,PhysRevLett.92.117902,PhysRevA.94.052311,PhysRevResearch.2.023165,PhysRevApplied.13.054079},
and the energy level of the SQUID qubits can be easily modulated by changing the local bias fields
\cite{RevModPhys.85.623}.
As shown in Fig.~\ref{squid}(a), the Hamiltonian of each SQUID qubit has the following form
\cite{PhysRevA.67.042311,PhysRevLett.92.117902}:
\begin{equation}
H_{SQUID}=\frac{Q^2}{2C}+\frac{(\mathbf{\Phi}-\mathbf{\Phi}_x)^2}{2L}-E_J\cos{(2\pi\frac{\mathbf{\Phi}}{\mathbf{\Phi}_0})},
\end{equation}
where $C$ is the junction capacitance,
$L$ is the loop inductance,
$Q$ is the total charge on the capacitor,
$\mathbf{\Phi}$ is the magnetic flux through the loop,
$\mathbf{\Phi}_x$ is the external flux applied to the loop,
$\mathbf{\Phi}_0=h/2e$ is the flux quantum,
and $E_J=I_c\mathbf{\Phi}_0/2\pi$ is the Josephson energy with $I_c$ being the critical current of the
junction.

Consider the system of four SQUID qubits coupled to a single-mode microwave cavity field, as shown in
Fig.~\ref{squid}(b).
Every SQUID qubit has a three-level structure with three states $\{|0\rangle_s, |1\rangle_s,|e\rangle_s\}
(s=1,2,3,4)$, as shown in Fig.~\ref{squid}(c), where $|0\rangle_s$ and $|1\rangle_s$ are two ground states
and $|e\rangle_s$ is the excited state.
The ground state $|0\rangle_s$ resonantly couples to the excited state $|e\rangle_s$ by the cavity field with the
constant strength $g_s$.
The classical field resonantly drives the transition between the states $|1\rangle_s$ and $|e\rangle_s$ with the
coupling strength  $\Omega_s$.
The general forms of coupling strength $\Omega_s$ and the coupling constant $g_s$ are given by
\cite{PhysRevA.67.042311,PhysRevLett.92.117902}
\begin{eqnarray}
\Omega_s&=&\frac{1}{2L_s\hbar}\langle1|\mathbf{\Phi}|e\rangle_s \int_{S_s} \mathbf{B}^s _{\mu
w}(\mathbf{r},t)\cdot d\mathbf{S}, \label{28a}\\
g_s&=&\frac{1}{L_s} \sqrt{\frac{\omega_c}{2\mu_0\hbar}}\langle0|\mathbf{\Phi}|e\rangle_s\int_{S_s}
\mathbf{B}^s _{c}(\mathbf{r})\cdot d\mathbf{S},   \label{28b}
\end{eqnarray}
where $S_s$ is the surface surrounded by the ring of the $s\mathrm{th}$ SQUID,
$L_s$ is the loop inductance of the $s\mathrm{th}$ SQUID,
$\omega_c$ is the angular frequency of the cavity, and $\mu_0$ is magnetic permeability in a vacuum.
$\mathbf{B}^s _{c}(\mathbf{r})$ is the magnetic component of the normal cavity mode.
$\mathbf{B}^s _{\mu w}(\mathbf{r},t)$ is the magnetic component of the classical field.

From Eqs.~(\ref{28a}) and (\ref{28b}), one finds that the coupling strength $\Omega_s$ and the coupling constant
$g_s$ can be controlled by the microwave magnetic components $\mathbf{B}^s _{\mu w}(\mathbf{r},t)$ and
the magnetic component $\mathbf{B}^s _{c}(\mathbf{r})$ of the normal cavity mode, respectively.
Due to the spatial inhomogeneity of the magnetic fields, there may exist uncertainty in $\Omega_s$ and $g_s$.
{On the other hand, the distorted pulse shape or inexact evolution time could also produce uncertainty in the pulse area. Consequently, the quantum operations would lose
accuracy in this system.}
To eliminate those uncertainties, we can employ the SMCPs scheme to prepare the $W$ state with high fidelity,
which is actually to control the microwave magnetic components $\mathbf{B}^s _{\mu w}(\mathbf{r},t)$, in practice.
In the following, we adopt $\eta_1$ and $\eta_2$ to represent the uncertainties from $\mathbf{B}^s _{\mu
w}(\mathbf{r},t)$ and $\mathbf{B}^s _{c}(\mathbf{r})$, respectively.
The Hamiltonian of the whole system in the interaction picture is $(\hbar=1)$ \cite{PhysRevA.94.052311}
\begin{eqnarray}
H_I&=&H_m+H_c, \label{27}\\
H_m&=&\frac{1}{2}\sum^4 _{s=1}\big[\Omega_s(1+\eta_1)|e\rangle_s\langle1|+\mathrm{H.c.}\big],
\\\label{28}
H_c&=&\frac{1}{2}\sum^4 _{s=1} \big[g_s(1+\eta_2)|e\rangle_s\langle0|\hat{a}+\mathrm{H.c.}\big],
\label{29}
\end{eqnarray}
where $\hat{a}$ is the annihilation operator of the cavity field.
$H_c$ ($H_m$) describes the interaction between the cavity (classical) fields and the SQUID qubits.

Note that the excited number is a conserved quantity in this system because the excited number operator
$\hat{N}_e=\sum_s(|e\rangle_s\langle e|+|1\rangle_s\langle 1|)+\hat{a}^{\dag}\hat{a}$ satisfies
$[\hat{N}_e,H_I]=0$.
Therefore, we can restrict the system dynamics in the single excited subspace, namely,
$\langle\psi(t)|\hat{N}_e|\psi(t)\rangle=1$, where $|\psi(t)\rangle$ is the evolution state of the system.
{The bases of the single excited subspace are}
\begin{eqnarray}  \label{33}
|\psi_1\rangle&=&|0\rangle_1|0\rangle_2|0\rangle_3|1\rangle_4|0\rangle_c,
~~~|\psi_2\rangle=|0\rangle_1|0\rangle_2|0\rangle_3|e\rangle_4|0\rangle_c,\nonumber\\[0.1ex]
|\psi_3\rangle&=&|0\rangle_1|0\rangle_2|0\rangle_3|0\rangle_4|1\rangle_c,
~~~|\psi_4\rangle=|e\rangle_1|0\rangle_2|0\rangle_3|0\rangle_4|0\rangle_c,\nonumber\\[0.1ex]
|\psi_5\rangle&=&|0\rangle_1|e\rangle_2|0\rangle_3|0\rangle_4|0\rangle_c,
~~~|\psi_6\rangle=|0\rangle_1|0\rangle_2|e\rangle_3|0\rangle_4|0\rangle_c,\nonumber\\[0.1ex]
|\psi_7\rangle&=&|1\rangle_1|0\rangle_2|0\rangle_3|0\rangle_4|0\rangle_c,
~~~|\psi_8\rangle=|0\rangle_1|1\rangle_2|0\rangle_3|0\rangle_4|0\rangle_c,\nonumber\\[0.1ex]
|\psi_9\rangle&=&|0\rangle_1|0\rangle_2|1\rangle_3|0\rangle_4|0\rangle_c. \nonumber
\end{eqnarray}
For simplicity,
we set $g_1=g_2=g_3=g$,
$g_4=\sqrt{3}g$. This could be realized by properly adjusting the parameters of the SQUIDs,
such as $L_s$ or $S_s$.
Then, we rewrite the Hamiltonian $H_c$ given by Eq.~(\ref{28}) in this set of basis, i.e.,
\begin{equation}
\begin{split}
H_c=&g(1+\eta_2)(|\psi_4\rangle+|\psi_5\rangle+|\psi_6\rangle)\langle\psi_3|\\
&+\sqrt{3}g(1+\eta_2)|\psi_2\rangle \langle\psi_3|+\mathrm{H.c.},
\end{split}
\end{equation}
and the eigenstates of the Hamiltonian $H_c$ are
\begin{eqnarray}
|\phi_1\rangle&=&-\frac{1}{\sqrt{2}}\Big[|\psi_2\rangle\!-
\!\frac{1}{\sqrt{3}}(|\psi_4\rangle\!+\!|\psi_5\rangle\!+\!|\psi_6\rangle)\Big],  \label{40a}\\
|\phi_2\rangle&=&\frac{1}{2}\Big[|\psi_2\rangle\!+
\!\sqrt{2}|\psi_3\rangle\!+\!\frac{1}{\sqrt{3}}(|\psi_4\rangle\!+\!|\psi_5\rangle\!+\!|\psi_6\rangle)\Big],
\label{40b}\\
|\phi_3\rangle&=&\frac{1}{2}\Big[|\psi_2\rangle\!-
\!\sqrt{2}|\psi_3\rangle\!+\!\frac{1}{\sqrt{3}}(|\psi_4\rangle\!+\!|\psi_5\rangle\!+\!|\psi_6\rangle)\Big],~~~~
\label{40c}
\end{eqnarray}
with the corresponding eigenvalues $E_1=0$, $E_2=\sqrt{6}g$, and $E_3=-\sqrt{6}g$. Remarkably,
$|\phi_1\rangle$ is the dark state of this system.
Next, we set $\Omega_1=\Omega_2=\Omega_3=\sqrt{2}\Omega_a$ and $\Omega_4=\sqrt{2}\Omega_b$, and the Hamiltonian $H_I$ of the whole system given by Eq.~(\ref{27}) can be rearranged into a new form,
\begin{eqnarray}
H'_I&=&H'_c+H'_m,  \label{38}\\
H'_c&=&\sum^3_{i=1}E_i|\phi_i\rangle\langle\phi_i|,\\
H'_m&=&\frac{\Omega_a}{\sqrt{2}}(1\!+\!\eta_1)\big(|\phi_2\rangle\!+\!|\phi_3\rangle\big)\langle\psi_1|
\!+\!\Omega_a(1\!+\!\eta_1)|\phi_1\rangle\langle\psi_1|\nonumber\\[0.8ex]
&&\!+\frac{\Omega_b}{\sqrt{2}}(1\!+\!\eta_1)\big(|\phi_2\rangle\!+\!|\phi_3\rangle\big)\langle
\psi_f|\!+\!\Omega_b(1\!+\!\eta_1)|\phi_1\rangle\langle \psi_f|\nonumber\\[0.8ex]
&&\!+\mathrm{H.c.},
\end{eqnarray}
where $|\psi_f\rangle=(|\psi_7\rangle+|\psi_8\rangle+|\psi_9\rangle)/\sqrt{3}$.
Finally, we perform the unitary transformation $U=e^{-i H'_ct}$ on this system, and
the Hamiltonian $H'_I$ given by Eq.~(\ref{38}) can be rewritten as
\begin{eqnarray}   \label{41}
H^{\prime\prime}_I&=&\frac{\Omega_a}{\sqrt{2}}(1\!+\!\eta_1)\Big(e^{i\sqrt{6}gt}|\phi_2\rangle+e^{-
i\sqrt{6}gt}|\phi_3\rangle\Big)\langle\psi_1|  \nonumber\\[0.8ex]
&&+\frac{\Omega_b}{\sqrt{2}}(1\!+\!\eta_1)\Big(e^{i\sqrt{6}gt}|\phi_2\rangle+e^{-i\sqrt{6}gt}|\phi_3\rangle\Big)\langle
\psi_f|  \nonumber\\[0.8ex]
&&+\Omega_a(1\!+\!\eta_1)|\phi_1\rangle\langle\psi_1|\!+\!\Omega_b(1\!+\!\eta_1)|\phi_1\rangle\langle
\psi_f|+\mathrm{H.c.}~~~~~~
\end{eqnarray}

\begin{figure}[b]
\centering
\scalebox{0.48}{\includegraphics{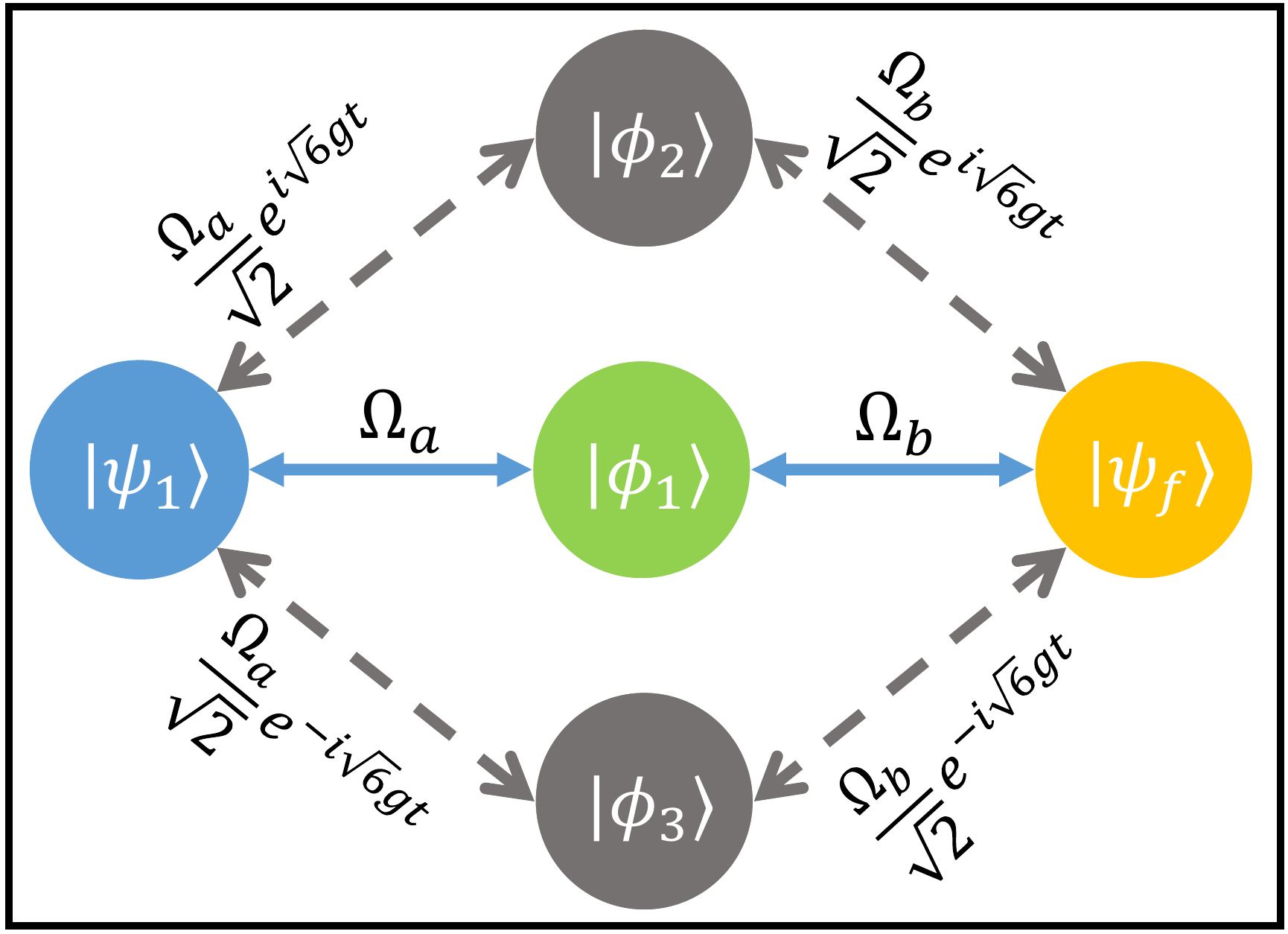}}
\caption{Schematic diagram of transition paths for the Hamiltonian $H^{\prime\prime}_I$ given by
Eq.~(\ref{41}).
}\label{paths}
\end{figure}

Figure~\ref{paths} shows the transition paths for the system governed by the Hamiltonian $H^{\prime\prime}_I$.
When the system parameters satisfy the condition $\Omega_{a,b}\ll g$,
the high-frequency oscillation terms $e^{\pm i\sqrt{6}gt}$ can be safely ignored.
Then, both the transition paths
$|\psi_1\rangle\!\!\leftrightarrow\!\!|\phi_2\rangle\!\!\leftrightarrow\!\!|\psi_f\rangle$ and
$|\psi_1\rangle\!\!\leftrightarrow\!\!|\phi_3\rangle\!\!\leftrightarrow\!\!|\psi_f\rangle$ are effectively suppressed, and
only the transition path $|\psi_1\rangle\!\!\leftrightarrow\!\!|\phi_1\rangle\!\!\leftrightarrow\!\!|\psi_f\rangle$ is feasible.
As a result, the whole system is simplified into a $\Lambda$-type physical model, and the effective Hamiltonian, in the basis $\{|\psi_1\rangle, |\psi_f\rangle, |\phi_1\rangle\}$, reads
\cite{PhysRevA.94.052311}
\begin{eqnarray}\label{Heq}
H_{e}=
(1\!+\!\eta_1)\left[
\begin{array}{ccc}	
0 &0& -\Omega_a \\
0 &0& \Omega_b \\
 -\Omega_a  & \Omega_b &0 \\
\end{array}	
\right],
\end{eqnarray}
where
$$|\psi_f\rangle\!=\!\frac{1}{\sqrt{3}}(|1\rangle_1|0\rangle_2|0\rangle_3+|0\rangle_1|1\rangle_2|0\rangle_3 +|0\rangle_1|0\rangle_2|1\rangle_3)\otimes|0\rangle_4|0\rangle_c$$
is the $W$ state.
Therefore, the SMCPs scheme studied in Sec.~\ref{cps} can be directly employed to the effective Hamiltonian given by Eq.~(\ref{Heq}) to prepare the $W$ state $|\psi_f\rangle$.
{Moreover,
we can expect that the influence caused by the uncertainty of the cavity field can be negligible during the evolution process because
$\eta_2$ does not appear in the effective Hamiltonian given by Eq.~(\ref{Heq}).}

\begin{figure}[b]
\centering
\scalebox{0.4}{\includegraphics{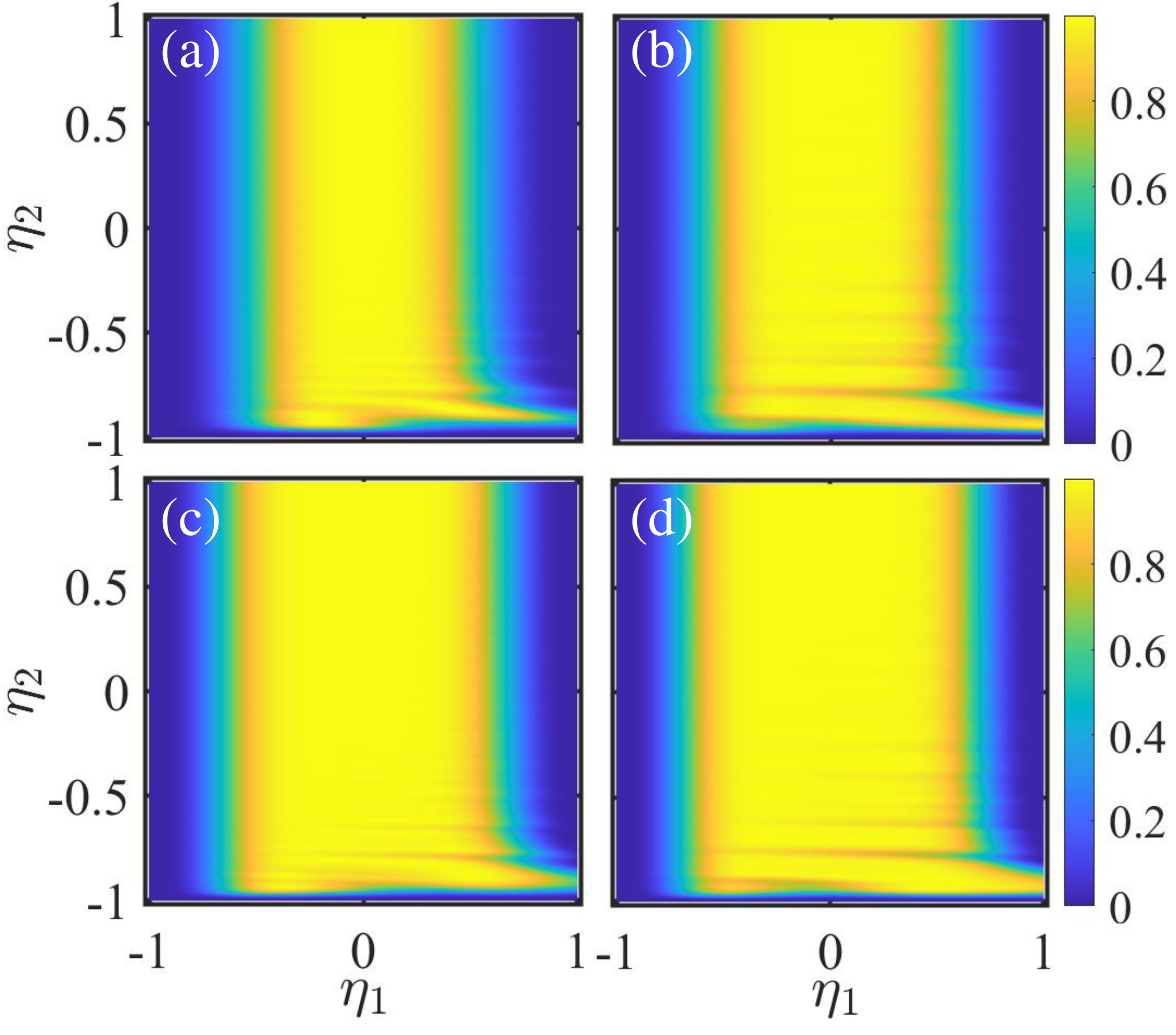}}
\caption{The fidelity $F$ of the $W$ state vs two kinds of the uncertainties $\eta_1$ and $\eta_2$ in the (a) two-pulses, (b) three-pulses, (c) four-pulses, and (d) five-pulses sequence. $g/\Omega_{max}=7$. $\tau=10^{3}$.
All $\Theta_n$ come from Table \ref{tab1}.
} \label{appl}
\end{figure}

In Figs.~\ref{appl}(a)--\ref{appl}(d),
we plot the fidelity of the $W$ state as a function of the uncertainties $\eta_1$ and $\eta_2$ with different pulses sequence,
where we set $\Omega_{max}=\max\{\Omega_{na},\Omega_{nb}\}$ and $g/\Omega_{max}=7$.
The numerical results verify that the uncertainty of the coupling constant $g$ has a negligible effect on
the final fidelity, since the fidelity almost remains unchanged with the increase of $\eta_2$.
Note that the fidelity obviously changes when $\eta_2<-0.5$. This is because the condition $g\gg \Omega_{a,b}$ is no longer satisfied very well. As a result, the effective Hamiltonian given by Eq.~(\ref{Heq}) is invalid in this regime.
On the other hand, the fidelity is also robust against the uncertainty of the coupling strength of the classical fields since the fidelity still keeps a relatively high value even though $\eta_1=\pm0.2$.
Thus, we can achieve the $W$ state in a robust way even when there are existing uncertainties in the superconducting circuits.

\begin{figure}[b]
\centering
\scalebox{0.57}{\includegraphics{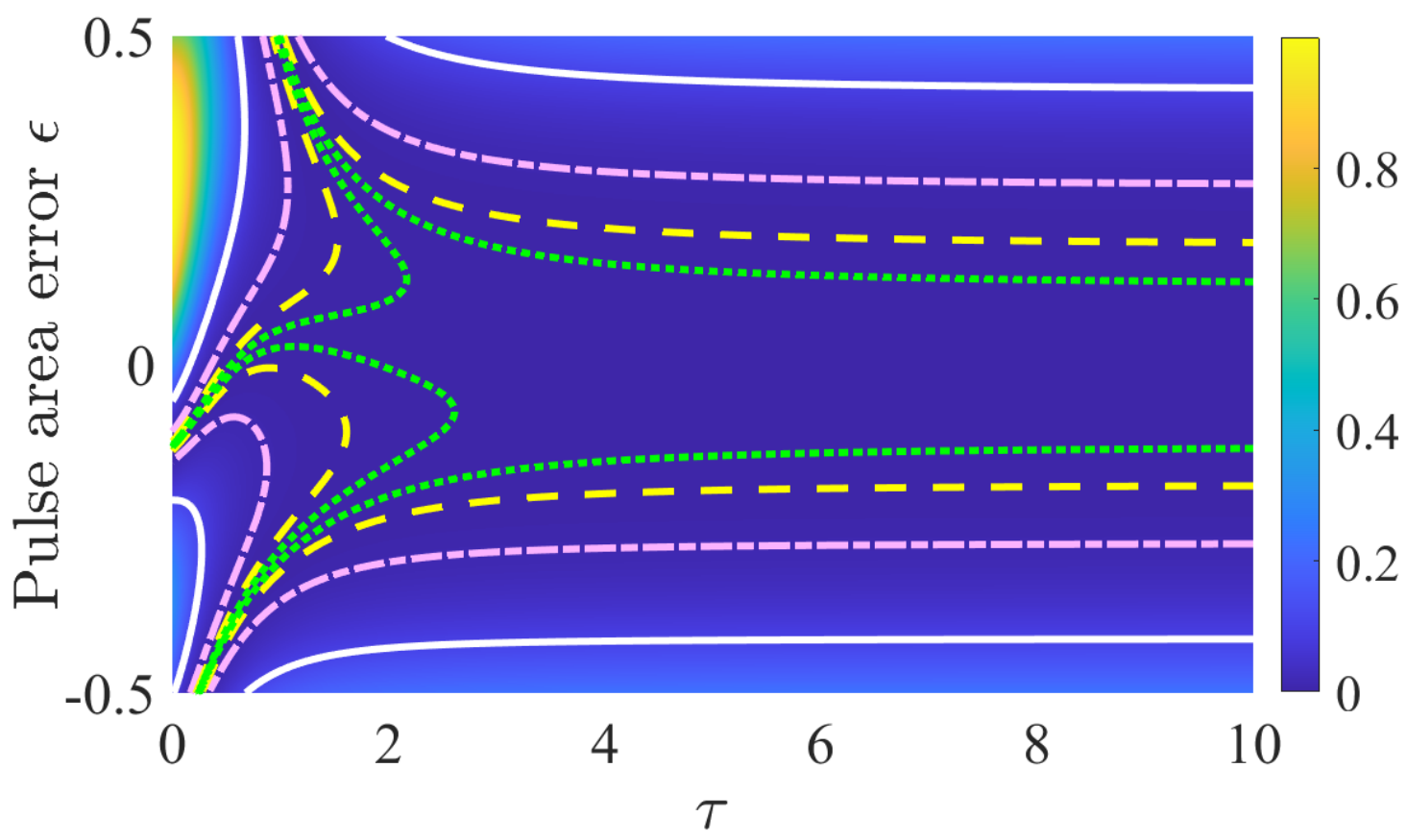}}
\caption{The infidelity $1-F$ of the $W$ state vs the uncertainty
$\eta_1$ and the deformation parameter $\tau$ in the three-pulses scheme, where $g/\Omega_{max}=15$. The green
dotted curve, the yellow dashed curve, the pink dot-dashed curve, and the white solid curve correspond to
the infidelity $1-F=10^{-4}$, $10^{-3}$, $10^{-2}$, and $10^{-1}$.
All $\Theta_n$ come from Table \ref{tab1}.}\label{appl2}
\end{figure}

Next, we turn to study the influence of waveform deformation on the fidelity of the $W$ state.
We take the three-pulses sequence as an example, and plot in Fig.~\ref{appl2} the infidelity as a function of the pulse area error and the waveform deformation.
An inspection of Fig.~\ref{appl2} demonstrates that the scheme is insensitive to waveform deformation, even in the case of severe deformation, because the $W$ state still maintains a high fidelity in the region $\tau>3$.
Hence, our scheme can resist the imperfection of input waveform and thus give more flexibility in terms of input waveform conditions.

Until now, we have not investigated the influence of decoherence on the fidelity of the $W$ state.
When considering the decoherence induced by the cavity decay,
the spontaneous emission, and the dephasing, the evolution of this system can be governed by the Lindblad
master equation,
\begin{equation}
\dot{\rho}=i[\rho,H_I]+\mathcal{L}(\sqrt{\kappa}\hat{a}) \rho
+\sum _{l=0}^{1}\sum^4 _{s=1}\Big[\mathcal{L}(\hat{\sigma}^s_{le})
\rho+\mathcal{L}(\hat{\sigma}^{s}_{ee,ll}) \rho\Big],\\
\end{equation}
where $\hat{\sigma}^s_{le}=\sqrt{\gamma_{ls}}|l \rangle_s\langle e|$,
$\hat{\sigma}^{s}_{ee,ll}=\sqrt{{\gamma^{\phi}_{ls}}/{2}}(|e \rangle_s\langle e|-|l \rangle_s\langle l|)$, and
the general form of the superoperator is
\begin{equation}
\mathcal{L}(\hat{o})
\rho=\hat{o}\rho\hat{o}^{\dag}-\frac{1}{2}(\hat{o}^{\dag}\hat{o}\rho+\rho\hat{o}^{\dag}\hat{o}).
\end{equation}
Here, $\hat{o}$ denotes the standard Lindblad operator,
$\kappa$ is the decay rate of the cavity field, and $\gamma_{ls}$ ($\gamma^{\phi}_{ls}$) represent the
dissipation (dephasing) rate from the excited state $|e\rangle$ to the ground state $|l\rangle$.

\begin{figure}[b]
\centering
\scalebox{0.36}{\includegraphics{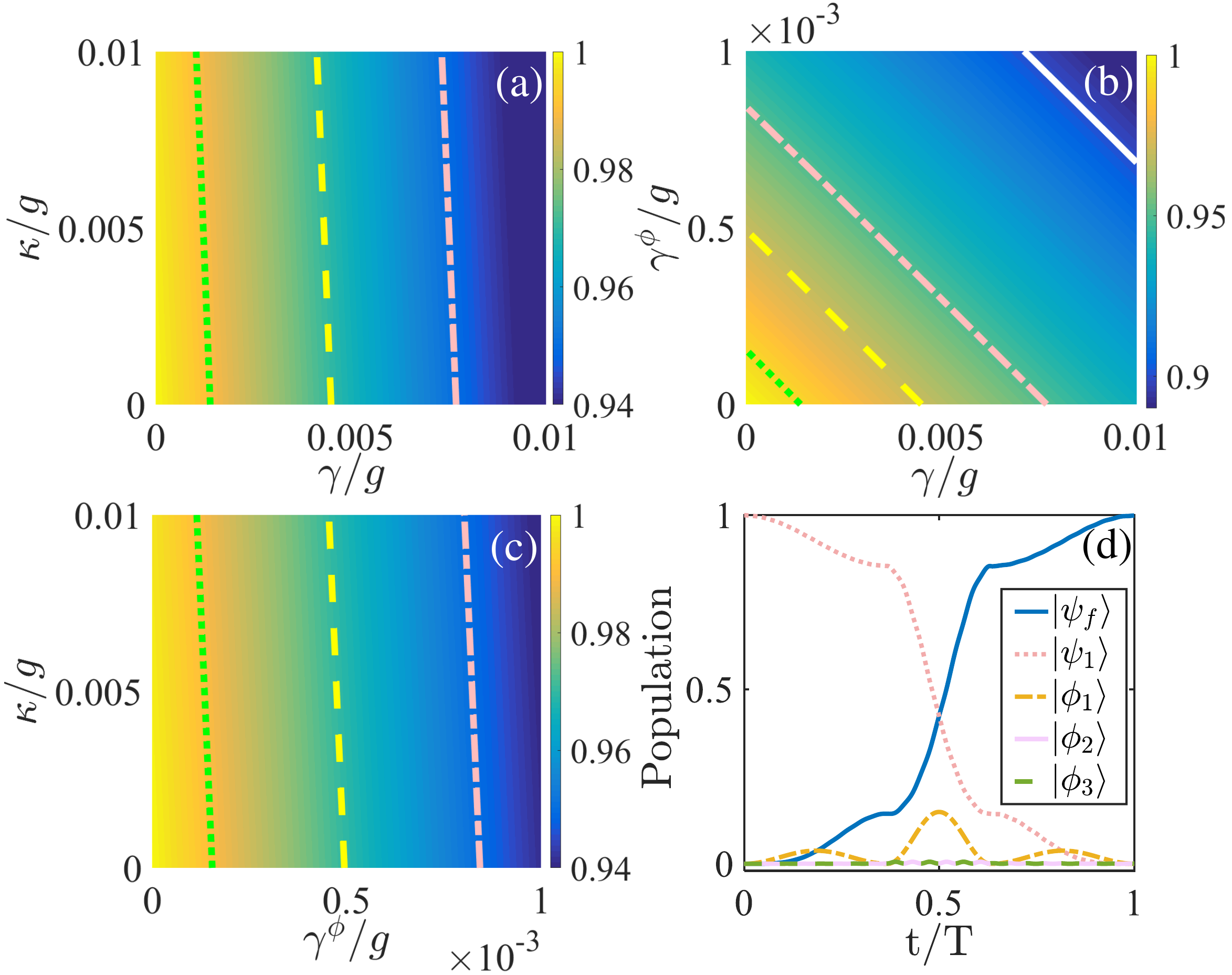}}
\caption{The fidelity of the $W$ state vs
(a) $\gamma$/g and $\kappa$/g,
(b) $\gamma$/g and $\gamma^{\phi}$/g,
and (c) $\gamma^{\phi}$/g and $\kappa$/g in the three-pulses scheme.
Here, the green dotted line, the yellow dashed line, the pink dot-dashed line, and the white solid line
correspond to the final fidelity $F=0.99$, $0.97$, $0.95$, and $0.90$. (d) The population evolution of the
states $|\phi_k\rangle$, $|\psi_1\rangle$, and $|\psi_f\rangle$ in the three pulses scheme, where the initial state
is $|\psi_1\rangle$.
$g/\Omega_{max}=7$ and $\tau=10^{3}$.
All $\Theta_n$ come from Table \ref{tab1}.}\label{deco}
\label{Fig.4}
\end{figure}

Figures~\ref{deco}(a)--\ref{deco}(c) demonstrate the relationship between the fidelity of the $W$ state and the decoherence
parameters $\{\gamma, \gamma^{\phi},\kappa\}$, where we set $\gamma_{ls}=\gamma$ and
$\gamma^{\phi}_{ls}=\gamma^{\phi}$ for simplicity.
According to Fig.~\ref{deco}(a) and Fig.~\ref{deco}(c),
the fidelity of the $W$ state is hardly affected by the cavity decay.
Physically,
it is due to the fact that the single-photon state of the cavity field is almost decoupling during the system evolution.
We can observe from Fig.~\ref{deco}(a) and Fig.~\ref{deco}(b) that the spontaneous emission of the SQUID
qubits has a slight influence on the fidelity of the $W$ state.
The reason is as follows.
Although $|\phi_k\rangle$ ($k=1,2,3$) contain the excited states of the SQUID qubits, as given by
Eqs.~(\ref{40a})--(\ref{40c}),
they are almost negligible in the evolution process when satisfying the condition $\Omega_{a,b}\ll g$. This can
be verified by Fig.~\ref{deco}(d), which demonstrates that the population of the state $|\phi_k\rangle$
($k=1,2,3$) is suppressed within an extremely small range during the system evolution.
As a result, the current scheme can maintain a high fidelity even when the dissipation rate of the excited
state is large.

{It is inevitable that the increase of the CPs sequence would prolong the total interaction time and impact the final fidelity under the decoherence environment.
In the following, we study this issue.
Figure~\ref{number}(a) shows the results obtained by various pulse lengths under different decoherence environments.
In the ideal situation,
i.e.,
$\gamma/g=0$,
the final fidelity almost remains unchanged as the pulse number increases.
When considering the decoherence caused by the atomic spontaneous emission and dephasing,
e.g.,
$\gamma/g=2\times10^{-5}$,
the final fidelity decreases slightly over an increasing number of pulses.
The reason is that a long pulse sequence takes a long interaction time so that the decoherence becomes the main factor in reducing the final fidelity.
Hence, to obtain a relatively high fidelity in the decoherence environment,
we cannot choose an overlong pulse sequence.

\begin{figure}[htbp]
\centering
\scalebox{0.36}{\includegraphics{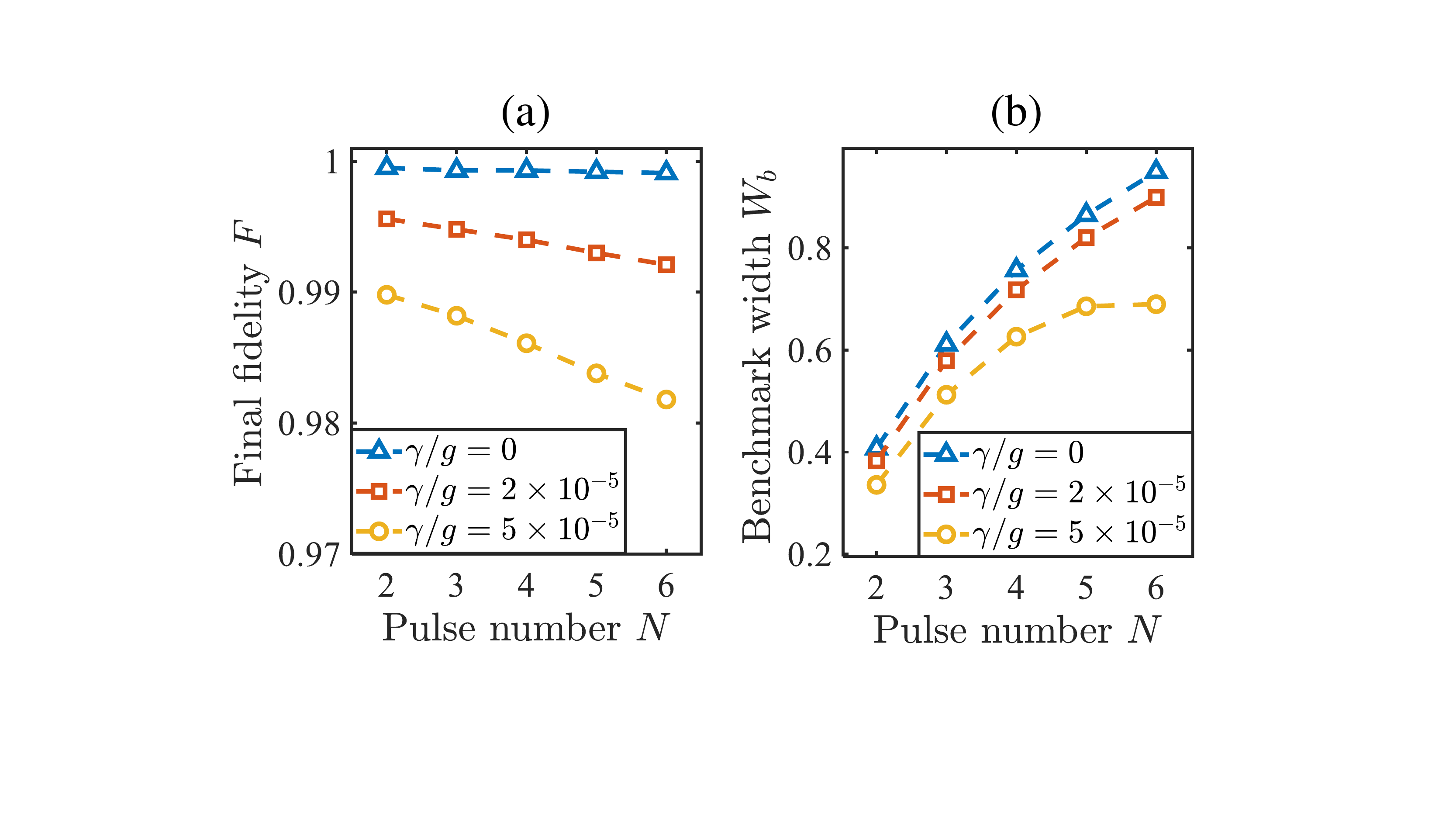}}
\caption{
{
(a) The final fidelity of the $W$ state and (b) the benchmark width $W_b$ vs different pulse number $N$.
Here we set the benchmark value $F_b=0.98$ and $\gamma_p=\gamma$.
All $\Theta_n$ come from Table \ref{tab1}.
Other parameters are
$\kappa/g=0.01$,
$g/\Omega_{max}=20$,
and $\tau=10^{3}$.
}}\label{number}
\end{figure}

To quantify the robustness under the decoherence environment,
we define a benchmark width
\begin{eqnarray}
W_b&=&\eta_{1,max}-\eta_{1,min},
\end{eqnarray}
where $\eta_{1,max(min)}$ is the positive (negative) maximum uncertainty in the coupling strength $\Omega_s$ to make the final fidelity reach a benchmark value $F_b$.
Remarkably,
this benchmark width can characterize the robustness against uncertainty provided by different pulse sequences.
Figure~\ref{number}(b) reveals the benchmark widths achieved by different pulse sequences under different decoherence environments.
On the one hand,
we can find that the too short pulse sequence cannot obtain a remarkable benchmark width (e.g.,
$W_b=0.3828$ for $N=2$ and $\gamma/g=2\times10^{-5}$).
On the other hand,
the benchmark width becomes wider along with the pulse number increasing,
and the increment of $W_b$ becomes gradually limited when the pulse number is large.
In other words,
when the system suffers from a serious decoherence,
we cannot obtain considerable robustness for an overlong sequence (e.g.,
$W_b=0.6899$ for $N=6$ and $\gamma/g=5\times10^{-5}$).
From this point of view,
to obtain excellent robustness,
both a too short pulse sequence and an overlong pulse sequence should not be considered preferentially.

Therefore,
we need to choose a moderate pulse length to seek an optimal tradeoff between the final fidelity and robustness (fault tolerance).
In recent experiments \cite{RevModPhys.85.623,PhysRevLett.113.123601,PhysRevLett.104.100504,Yan2016,PhysRevLett.125.180503,PhysRevLett.124.120501},
the following parameters are feasible:
$\gamma=\gamma^{\phi} \sim 10\mathrm{kHz}$ and $\kappa \sim625\mathrm{kHz}$.
The coupling constant $g$ could reach $600\mathrm{MHz}$ \cite{Niemczyk2010},
and thus $\gamma/g\sim1.6667\times10^{-5}$.
According to these parameters,
the final fidelity of the $W$ state can reach 0.9955 by the four-pulses sequence,
while the benchmark width is $W_b=0.7307$ for $F_b=0.98$.
As a result,
our scheme still retains good performance in a decoherence environment.}

\section{Conclusion}\label{con}

In summary,
we have put forward the SMCPs scheme to achieve robust population inversion in a three-level system.
The transition probability is directly derived from a total propagator, and we mainly focus on reducing the influence caused by the pulse area error on the transition probability.
By means of the Taylor expansion, we nullify the first few error terms in the transition probability through designing the values of the coupling strengths.
We have given the analytical expressions of the coupling strengths in the two-pulses sequence. For the three- and more-than-three-pulses (up to seven-pulses) sequence, we present a set of optimal parameters by numerical methods.
These results indicate that the robust manner against the pulse area error can be much better as the pulse number increases.
Our scheme provides a feasibility extension of a longer sequence.
Meanwhile, this scheme provides a flexible choice to design pulse length in a desirable manner for achieving high robustness against errors and an appropriate total operating time.
{Moreover, the pulse sequence interfered by waveform deformation is still valid to achieve high transition probability.}
This property is very useful for the resistance of pulse jump delay.
As a result,
our SMCPs scheme is robust against the pulse area error and waveform deformations.

Then, we have extended this SMCPs scheme to prepare the $W$ state with high fidelity in the superconducting circuits,
where four SQUID qubits are coupled with the cavity field and classical fields.
The SMCPs that we design can effectively compensate for the adverse impact caused by waveform deformations and uncertainties in coupling strengths.
Furthermore, we also discuss the feasibility and the robustness of this scheme in a decoherence environment.
The simulation results show that our scheme can obtain a good performance against the impact of the cavity decay and the spontaneous emission,
{and we need to select a moderate pulse length to achieve robust and accurate quantum control.}
We hope that this scheme will provide robust control for more physical models in the future.

\begin{acknowledgments}
	This work is supported by the National Natural Science Foundation of China under Grants No.~11874114 and No.~11805036, the Natural Science Foundation of Fujian Province under Grant No.~2021J01575, the Natural Science Funds for Distinguished Young Scholar of Fujian Province under Grant No. 2020J06011, and the Project from Fuzhou University under Grant No. JG202001-2.
\end{acknowledgments}

\begin{appendix}

\begin{widetext}

{
\section{Derivation of the propagator for the $N$ pulses sequence in the two-photon resonance regime} \label{A1}
In this appendix,
we deduce the expression of the propagator without errors when the three-level system works in the two-photon resonance regime,
where all parameters are time-independent.
The general form of a three-level system Hamiltonian for the $n\mathrm{th}$ pulse is given by
\begin{equation}  \label{11}
H_n=\Delta_n|e\rangle\langle e|+\frac{1}{2} \Big(\Omega_n e^{i\phi_n}|g\rangle\langle e|+\lambda_n e^{i
\varphi_n}|f\rangle\langle e|+\mathrm{H.c.}\Big),
\end{equation}
where $\Omega_n$ and $\phi_n$ ($\lambda_n$ and $\varphi_n$) are the coupling strength and the phase of the transition
$|g\rangle\leftrightarrow|e\rangle $ ($|f\rangle\leftrightarrow|e\rangle$), and $\Delta_n$ is the single-photon detuning.
In this case,
the propagator becomes $U_n(\Theta_n)=\exp\big({-i H_n T_n}\big)$,
where $T_n$ is the pulse duration and $\Theta_n=\arctan(\Omega_n/\lambda_n)$.
By selecting the pulse duration $T_n=2\pi/\sqrt{\Omega_n^2+\lambda_n^2+\Delta_n^2}$,
two ground states are decoupled from the excited state $|e\rangle$ so that we only focus on the subspace $\{|g\rangle,|f\rangle\}$.
Thus, the matrix form of the propagator $U_n(\Theta_n)$ in the basis $\{|g\rangle,|f\rangle\}$ reads
\begin{eqnarray}
U_n(\Theta_n)
&=&
\cos{x_n}\left[
\begin{array}{cc}\label{propaga}	
\cos{2\Theta_n}+i \tan{x_n}&-e^{i \Psi_n}\sin{2\Theta_n}  \\
-e^{-i \Psi_n}\sin{2\Theta_n} &-\cos{2\Theta_n}+i \tan{x_n} \\
\end{array}	
\right],
\end{eqnarray}
where the global phase is neglected, and
\begin{eqnarray}
x_n=\frac{\pi\Delta_n}{2\sqrt{1+\Delta^2_n}},~~~~\Psi_n&=&\phi_n-\varphi_n.\nonumber
\end{eqnarray}

Note that the total propagator for the composite pulses sequence could be obtained by nesting the propagator $U_n(\Theta_n)$ in order.
We first start from the two-pulses sequence, and the propagator comes into
\begin{eqnarray}
U^{(2)}&=&U_2(\Theta_2)U_1(\Theta_1)\nonumber\\
&=&
\cos{x_1}\cos{x_2}\left[
\begin{array}{cc}	
\cos{2\Theta_2}+i \tan{x_2}&-e^{i \Psi}\sin{2\Theta_2}  \\
-e^{-i \Psi}\sin{2\Theta_2} &-\cos{2\Theta_2}+i \tan{x_2} \\
\end{array}	
\right]
\left[
\begin{array}{cc}	
\cos{2\Theta_1}+i \tan{x_1}&-e^{i \Psi}\sin{2\Theta_1}  \\
-e^{-i \Psi}\sin{2\Theta_1} &-\cos{2\Theta_1}+i \tan{x_1} \\
\end{array}	
\right]\nonumber\\
&=&
\cos{x_1}\cos{x_2}
\left[
\begin{array}{cc}	
\cos{2(\Theta_1-\Theta_2)}+g_2&\big[-\sin{2(\Theta_1-\Theta_2)}+h_2\big]e^{i \Psi}  \nonumber\\[.3ex]
\big[\sin{2(\Theta_1-\Theta_2)}-h_2^*\big]e^{-i \Psi}&-\cos{2(\Theta_1-\Theta_2)}+g_2^*  \nonumber \\[.3ex]
\end{array}	
\right],
\end{eqnarray}
with
\begin{eqnarray}
g_2&=&-\tan{x_1}\tan{x_2}+i\left(\tan{x_1}\cos{2\Theta_2}+\tan{x_2}\cos{2\Theta_1}\right),\nonumber\\
h_2&=&-i\left(\tan{x_1}\sin{2\Theta_2}+\tan{x_2}\sin{2\Theta_1}\right),\nonumber
\end{eqnarray}
where $g_2$ and $h_2$ are defined as functions associated with the parameters $\{x_1,x_2,\Theta_1,\Theta_2\}$.
Similarly, the propagator of the three-pulses sequence is
\begin{eqnarray}
U^{(3)}&=&U_3(\Theta_2)U_2(\Theta_2)U_1(\Theta_1)\nonumber\\
&=&
\left(\prod_{n=1}^{3}\cos{x_n}\right)
\left[
\begin{array}{cc}	
\cos{2(\Theta_1-\Theta_2+\Theta_3)+g_3}&\big[-\sin{2(\Theta_1-\Theta_2+\Theta_3)}+h_3\big]e^{i \Psi}  \nonumber\\[.3ex]
\big[-\sin{2(\Theta_1-\Theta_2+\Theta_3)}+h_3^*\big]e^{-i \Psi}&-\cos{2(\Theta_1-\Theta_2+\Theta_3)}-g_3^*  \nonumber
\end{array}	
\right],
\end{eqnarray}
where
\begin{eqnarray}
g_3&=&-(\tan{x_{1}}\tan{x_{2}}\cos{2\Theta_3}+\tan{x_{1}}\tan{x_{3}}\cos{2\Theta_2}+\tan{x_{2}}\tan{x_{3}}\cos{2\Theta_1})\nonumber\\
&&+i\big[\tan{x_{1}}\cos{2(\Theta_2-\Theta_3)}+\tan{x_{2}}\cos{2(\Theta_1-\Theta_3)}+\tan{x_{3}}\cos{2(\Theta_1-\Theta_2)}-\tan{x_1}\tan{x_2}\tan{x_3}\big],\nonumber\\
h_3&=&(\tan{x_{1}}\tan{x_{2}}\sin{2\Theta_3}+\tan{x_{1}}\tan{x_{3}}\sin{2\Theta_2}+\tan{x_{2}}\tan{x_{3}}\sin{2\Theta_1})\nonumber\\
&&-i\big[\tan{x_{1}}\sin{2(\Theta_2-\Theta_3)}+\tan{x_{2}}\sin{2(\Theta_1-\Theta_3)}+\tan{x_{3}}\sin{2(\Theta_1-\Theta_2)}\big].\nonumber
\end{eqnarray}
Following the above rules,
the propagator of the $N$-pulses sequence could be obtained and reads
\begin{eqnarray}  \label{A3}
U^{(N)}&=& U_N(\Theta_N)U_{N-1}(\Theta_{N-1})\cdots U_1(\Theta_1) \nonumber \\
&=&
\left(\prod_{n=1}^{N}\cos{x_n}\right)\left[
\begin{array}{cc}	
\cos{2\Theta_N}+i \tan{x_N}&-e^{i \Psi}\sin{2\Theta_N}  \\
-e^{-i \Psi}\sin{2\Theta_N} &-\cos{2\Theta_N}+i \tan{x_N} \\
\end{array}	
\right]
\cdots
\left[
\begin{array}{cc}	
\cos{2\Theta_1}+i \tan{x_1}&-e^{i \Psi}\sin{2\Theta_1}  \\
-e^{-i \Psi}\sin{2\Theta_1} &-\cos{2\Theta_1}+i \tan{x_1} \\
\end{array}	
\right]\nonumber\\
&=&
\left(\prod_{n=1}^{N}\cos{x_n}\right)\left[
\begin{array}{cc}	
\cos{(2\sum^{N}_{n=1}(-1)^{n+1}\Theta_n)}+g_N&\left[-\sin{(2\sum^{N}_{n=1}(-1)^{n+1}\Theta_n)}+h_N\right]e^{i \Psi}  \\[.3ex]
(-1)^{N}\left[\sin{(2\sum^{N}_{n=1}(-1)^{n+1}\Theta_n)}-h_N ^* \right] e^{-i \Psi} &(-1)^{N}\left[\cos{(2\sum^{N}_{n=1}(-1)^{n+1}\Theta_n)}+ g_N ^*\right]
\end{array}	
\right]\!,~~~
\end{eqnarray}
where
\begin{eqnarray}
g_N&=&\sum^{N-1}_{a=1}\left[\prod^N_{1\leq j_1\leq\cdots \leq j_a\leq N} \left(i \tan{x_{j_l}} \right) \right] \cos{\left[ \sum^N_{\substack{1\leq j'_1\leq\cdots \leq j'_{N-a}\leq N\\j_l\neq j'_m}}(-1)^{(m+1)}\left(2\Theta_{j'_m}\right) \right]}+\prod^N_{n=1}\left(i \tan{x_n}\right),\nonumber\\
h_N&=&-\sum^{N-1}_{a=1}\left[\prod^N_{1\leq j_1\leq\cdots \leq j_a\leq N} \left(i \tan{x_{j_l}} \right) \right] \sin{\left[ \sum^N_{\substack{1\leq j'_1\leq\cdots \leq j'_{N-a}\leq N\\j_l\neq j'_m}}(-1)^{(m+1)}\left(2\Theta_{j'_m}\right) \right]}.\nonumber
\end{eqnarray}

We observe from Eq.~(\ref{A3}) that the transition probability of the target state $|f\rangle$ is concurrently controlled by the coupling strengths and detuning.
When the system is always in the resonance regime ($\Delta_n=0$),
it is easily found that $\cos{x_n}=1$, $g_{N}=0$, and $h_{N}=0$.
Then, the final propagator in the absence of errors becomes very concise, which reads
\begin{eqnarray}
U^{(N)}=
\left[
\begin{array}{cc}	
\cos{\left[2\sum^{N}_{n=1}(-1)^{n+1}\Theta_n\right]}&-\sin{\left[2\sum^{N}_{n=1}(-1)^{n+1}\Theta_n\right]}e^{i \Psi}  \nonumber\\[.3ex]
(-1)^{N}\sin{\left[2\sum^{N}_{n=1}(-1)^{n+1}\Theta_n\right]}e^{-i \Psi}&(-1)^N\cos{\left[2\sum^{N}_{n=1}(-1)^{n+1}\Theta_n\right]}  \nonumber \\[.3ex]
\end{array}	
\right].
\end{eqnarray}
As a result, the expression of the transition probability of the target state $|f\rangle$ is
\begin{eqnarray}
P_N =|U_{21}|^2&=&\sin^2{\left[2\sum^{N}_{n=1}(-1)^{n+1}\Theta_n\right]}, \nonumber
\end{eqnarray}
which is Eq.~(\ref{28a}) in the main text.
In order to achieve complete population inversion,
the ratio of coupling strength for each pulse, obtained by solving the equation $P_N =1$, must be satisfied, and the solution is
\begin{eqnarray}
\sum^{N}_{n=1}(-1)^{n+1}\Theta_n=\frac{\pi}{4}+\frac{m\pi}{2},
\end{eqnarray}
where $m$ is an arbitrary integer. This is the first condition that $\Theta_n$ should be satisfied.}

\section{Detailed derivation process in the two pulses sequence} \label{AAA}

In this appendix,
we present the detailed derivation process of the coupling strengths in the two-pulses sequence. When $\Delta=0$, according to Eq.~(\ref{ures}) in the main text, the total propagator in the  basis $\{|g\rangle,|f\rangle,|e\rangle\}$ is given by
\begin{eqnarray}
U^{(2)}&=&U_2(A_2,\Theta_2,\phi_2,\varphi_2)U_1(A_1,\Theta_1,\phi_1,\varphi_1)\notag\\[.8ex]
&=&\left[\begin{array}{ccc}	
\displaystyle\cos^2{\Theta_2}+\sin^2{\Theta_2}\cos{\frac{A_2}{2}} &\displaystyle -e^{i
\Psi_2}\sin^2{\frac{A_2}{4}}\sin{2 \Theta_2}  &\displaystyle -i e^{i \phi_2}\sin{\Theta_2}\sin{\frac{A_2}{2}} \\[1.5ex]
\displaystyle-e^{-i \Psi_2}\sin^2{\frac{A_2}{4}}\sin{2 \Theta_2} &\displaystyle
\sin^2{\Theta_2}+\cos^2{\Theta_2}\cos{\frac{A_2}{2}} &\displaystyle -i e^{i
\varphi_2}\cos{\Theta_2}\sin{\frac{A_2}{2}} \\[1.5ex]
\displaystyle-i e^{-i \phi_2}\sin{\Theta_2}\sin{\frac{A_2}{2}} &\displaystyle -i e^{-i \varphi_2}\cos{\Theta_2}\sin{\frac{A_2}{2}}
&\displaystyle \cos{\frac{A_2}{2}} \\[1.5ex]
\end{array}	
\right]\notag\\[.8ex]
&&\times\left[\begin{array}{ccc}	
\displaystyle\cos^2{\Theta_1}+\sin^2{\Theta_1}\cos{\frac{A_1}{2}} &\displaystyle -e^{i
\Psi_1}\sin^2{\frac{A_1}{4}}\sin{2 \Theta_1}  &\displaystyle -i e^{i \phi_1}\sin{\Theta_1}\sin{\frac{A_1}{2}} \\[1.5ex]
\displaystyle-e^{-i \Psi_1}\sin^2{\frac{A_1
}{4}}\sin{2 \Theta_1} &\displaystyle
\sin^2{\Theta_1}+\cos^2{\Theta_1}\cos{\frac{A_1}{2}} &\displaystyle -i e^{i
\varphi_1}\cos{\Theta_1}\sin{\frac{A_1}{2}} \\[1.5ex]
\displaystyle-i e^{-i \phi_1}\sin{\Theta_1}\sin{\frac{A_1}{2}} &\displaystyle -i e^{-i \varphi_1}\cos{\Theta_1}\sin{\frac{A_1}{2}}
&\displaystyle \cos{\frac{A_1}{2}} \\[1.5ex]
\end{array}	
\right]\notag\\[.8ex]
&=&\left[\begin{array}{ccc}	
U^{(2)}_{11}&U^{(2)}_{12}&U^{(2)}_{13}\\[.8ex]
U^{(2)}_{21}&U^{(2)}_{22}&U^{(2)}_{23}\\[.8ex]
U^{(2)}_{31}&U^{(2)}_{32}&U^{(2)}_{33}\\[.8ex]
\end{array}	
\right].
\end{eqnarray}
When the system is in the ground state $|g\rangle$,
the time evolution of the system state reads $|\psi(t)\rangle=U^{(2)}_{11}|g\rangle+U^{(2)}_{21}|f\rangle+U^{(2)}_{31}|e\rangle$.
Consider the error $\epsilon$ occurring in the pulse area: $\tilde{A}_n=A_n(1+\epsilon)$, $n=1,2$.
The transition amplitude of the state $|f\rangle$ can be rewritten as
\begin{eqnarray}\label{a2}
U^{(2)}_{21}(\epsilon)&=&-e^{-i\Psi_1}\sin^2{\frac{\tilde{A}_1}{4}}\sin{2\Theta_1}(\sin^2{\Theta_2}+\cos{\frac{\tilde{A}_2}{2}}\cos^2{\Theta_2})-e^{-i\Psi_2}\sin^2{\frac{\tilde{A}_2}{4}}\sin{2\Theta_2}(\cos^2{\Theta_1}+\cos{\frac{\tilde{A}_1}{2}}\sin^2{\Theta_1})\cr
&&-e^{-i(\phi_1-\varphi_2)}\sin{\Theta_1}\cos{\Theta_2}\sin{\frac{\tilde{A}_1}{2}}\sin{\frac{\tilde{A}_2}{2}},
\end{eqnarray}
where $\Psi_n=\phi_{n}-\varphi_{n}$.

The transition probability of the state $|f\rangle$ is equal to the modular squaring of the probability amplitude,
i.e.,
$|U^{(2)}_{21}|^2$.
Hence,
the transition probability carries the erroneous pulse area $\tilde{A}_n$ as
\begin{eqnarray}  \label{a4}
P_2(\epsilon)&=&\sin ^2\!\Theta_1 \!\cos ^2\!\Theta_2 \!\sin ^2\!\frac{\tilde{A}_1}{2} \!\sin ^2\!\frac{\tilde{A}_2}{2}\!+\!8 \big[ \cos ^2\!\Theta_2 \!\sin \!\Theta_1 \!\sin \!\Theta_2 \!\sin ^3\!\frac{\tilde{A}_2}{4} \!\sin \!\frac{\tilde{A}_1}{2} \!\cos \!\frac{\tilde{A}_2 }{4}\!\cos (\phi_1\!-\!\phi_2)\cos \!2\Theta_1 \sin ^2\!\frac{\tilde{A}_1}{4}\!+\!\cos ^2\!\frac{\tilde{A}_1}{4}\nonumber\\[.1ex]
&&+\sin ^2\!\Theta_1\!\cos\!\Theta_1\!\cos\!\Theta_2\!\sin ^3\!\frac{\tilde{A}_1}{4}  \cos \!\frac{\tilde{A}_1}{4}  \sin \!\frac{\tilde{A}_2}{2} \! \cos (\varphi_1\!-\!\varphi_2) \cos ^2\Theta_2 \!\cos \!\frac{\tilde{A}_2}{2} \!+\!\sin ^2\!\Theta_2\big]\!\nonumber\\[.1ex]
&&+\!\sin ^2\!2\Theta_1\!\sin ^4\frac{\tilde{A}_1}{4}\Big[\cos ^2\Theta_2 \cos \frac{\tilde{A}_2}{2}\!+\!\sin ^2\!\Theta_2\Big]^2+\sin ^2\!2\Theta_2 \sin ^4\frac{\tilde{A}_2}{4}\Big[\sin ^2\Theta_1 \cos \frac{\tilde{A}_1}{2}\!+\!\cos ^2\Theta_1\Big]^2\!\nonumber\\[.1ex]
&&+2 \sin 2\Theta_1 \sin 2\Theta_2 \sin ^2\!\frac{\tilde{A}_1}{4}  \!\sin ^2\frac{\tilde{A}_2}{4}\!\Big[\cos 2\Theta_1 \sin ^2\frac{\tilde{A}_1}{4} \!+\!\cos ^2\frac{\tilde{A}_1}{4} \Big] \Big[\cos ^2\Theta_2 \cos \frac{\tilde{A}_2}{2}+\sin ^2\Theta_2\Big] \!\cos \!\Gamma_1,
\end{eqnarray}
where $\Gamma_1=\Psi_1-\Psi_2$.
After substituting the pulse area $A_1=A_2=2\pi$ into Eq.~(\ref{a4}), we have
\begin{eqnarray}
P_2(\epsilon) &=&\sin ^2\Theta_1 \cos ^2\Theta_2 \sin ^4\pi  \epsilon\!+\!\cos^4 \frac{\pi  \epsilon}{2}\big[\sin ^22\Theta_1  (\sin ^2\Theta_2\!-\!\cos ^2\Theta_2 \cos \pi  \epsilon )^2\!+\!\sin ^22\Theta_2(\cos ^2\Theta_1\!-\!\sin ^2\Theta_1 \cos \pi  \epsilon )^2\big]\nonumber\\[.1ex]
&&+2 \sin 2\Theta_1 \sin 2\Theta_2 \cos ^4\frac{\pi  \epsilon }{2}(\cos 2\Theta_1 \cos ^2\frac{\pi  \epsilon }{2}+\sin ^2\frac{\pi  \epsilon }{2}) (\sin ^2\Theta_2-\cos ^2\Theta_2 \cos \pi  \epsilon) \cos\Gamma_1\nonumber\\[.1ex]
&&+4\sin^2 \pi \epsilon  \cos ^2\!\frac{\pi  \epsilon }{2} \big[\sin \Theta_1 \sin \Theta_2 \cos ^2\Theta_2 \cos (\phi_1-\phi_2) (\cos 2\Theta_1 \cos ^2\frac{\pi \epsilon }{2}+\sin ^2\frac{\pi  \epsilon }{2})\nonumber\\[.1ex]
&&+\cos \Theta_1\cos \Theta_2\sin ^2\Theta_1\cos (\varphi_1-\varphi_2) (\sin ^2\Theta_2-\cos ^2\Theta_2 \cos \pi  \epsilon)\big].
\end{eqnarray}

Then, by the Taylor expansion,
$P_2(\epsilon)$ could be written as the following form (up to the second-order term):
\begin{equation}
P_2(\epsilon)=\alpha_{2,0}+\alpha_{2,2}\epsilon^2+O(\epsilon^4).
\end{equation}
Here, the zero-order is the precise transition probability, and the goal of designing the composite pulses sequence is to eliminate as many of the high-order coefficients as possible. Note that all odd-order coefficients are eliminated due to the choice of $A_1=A_2=2\pi$. The expressions of the zeroth-order and the second-order coefficients are
\begin{eqnarray}
\alpha_{2,0}&=&\frac{1}{2}\big(1-\cos {4\Theta_1}\cos {4\Theta_2 }-\cos{\Gamma_1}\sin{4 \Theta_1}\sin{4 \Theta_2}\big),\label{a7}\\[.8ex]
\alpha_{2,2}&=&\frac{\pi^2}{4}\big[\sin\!4\Theta_1(\sin\!2\Theta_2\!+\!2\!\sin\!4\Theta_2)\!-\!\sin\!2\Theta_1\!\sin\!4\Theta_2\big]\!\cos\!\Gamma_1\!+\!\frac{\pi^2}{2}(\cos\!4\Theta_1\!\cos\!4\Theta_2\!-\!\sin^2\!2\Theta_1\!\cos\!2\Theta_2\!+\!\cos\!2\Theta_1\!\sin^2\!2\Theta_2\!-\!1)\cr
&&+2\pi^2\big[\sin\Theta_2\cos^2\Theta_2(\sin3\Theta_1-\sin\Theta_1)\cos(\phi_1-\phi_2)-\sin^2\Theta_1\cos\Theta_1(\cos\Theta_2+\cos3\Theta_2)\cos(\varphi_1-\varphi_2)\big]
 \label{a8}
\end{eqnarray}

In the two-pulses sequence, there are only two parameters, $\Theta_1$ and $\Theta_2$. So, only two equations can be satisfied, i.e.,
\begin{numcases}{}
\alpha_{2,0}=1, \label{a9}\\
\alpha_{2,2}=0 \label{a10}.
\end{numcases}
Note that the expression given by Eq.~(\ref{a7})
can be rearranged by $\alpha_{2,0}=1/2\big[1-\mathcal{A}\cos {(4\Theta_1-\beta)}\big]$, where $\mathcal{A}=\sqrt{\cos^2{4\Theta_2}+\cos^2{\Gamma_1}\sin^2{4\Theta_2}}$.
Remarkably, the following inequality is always satisfied:
\begin{eqnarray}\label{A7}
\mathcal{A}=\sqrt{\cos^2{4\Theta_2}+\cos^2{\Gamma_1}\sin^2{4\Theta_2}}\leq\sqrt{\cos^2{4\Theta_2}+\sin^2{4\Theta_2}}=1.
\end{eqnarray}
Only when $\Gamma_1=m\pi$ could there be an equality, where $m$ is an arbitrary integer. In other words, the equation $\alpha_{2,0}=1$ has real solutions if and only if $\Gamma_1=m\pi$. For simplicity, we set $\Gamma_1=0$ in the main text, and Eqs.~(\ref{a9}) and (\ref{a10}) become
\begin{numcases}{}
\sin^22(\Theta_1-\Theta_2 )=1,\label{two11}\\[.8ex]
\pi ^2 \sin 2 (\Theta_2-\Theta_1) \Big[\sin 2 (\Theta_1-\Theta_2)
\!+\!\cos \Theta_2 (2 \sin \Theta_1\!+\!\sin \Theta_2)\!+\!\sin \Theta_1 \cos \Theta_1\Big]=0\label{two22}.
\end{numcases}
One solution of Eqs.~(\ref{two11}) and (\ref{two22}) can be written as
\begin{equation} \label{PHIE}
\begin{split}
\Theta_1=m\pi+\frac{\pi}{8}, ~\Theta_2=m\pi+\frac{3\pi}{8},
\end{split}
\end{equation}
where $m$ is an arbitrary integer.
With this group of solutions,
the final propagator without errors becomes (in the basis $\{|g\rangle, |f\rangle,|e\rangle\}$)
\begin{eqnarray}
U^{(2)}=
i\left[
\begin{array}{ccc}	
0&1 &0 \\
1&0 &0 \\
0&0 &-i \\
\end{array}	
\right].
\end{eqnarray}
This propagator is recognized as an $X$ gate with the global phase factor $i$.
For the initial state of the system $|\psi_i\rangle=|g\rangle$,
the final state would become $|\psi_f\rangle=i|f\rangle$.
Similarly,
when the phase difference $\Psi=0$,
by this group of solutions,
we could also obtain a $Y$ gate,
\begin{eqnarray}
U^{(2)}=
i\left[
\begin{array}{ccc}	
0&-i &0 \\
i&0 &0 \\
0&0 &-i \\
\end{array}	
\right]
\end{eqnarray}
with the same global phase factor.
For the initial state of the system $|\psi_i\rangle=|g\rangle$,
the final state would become $|\psi_f\rangle=-|f\rangle$.
Note that both the $X$ gate and the $Y$ gate could achieve the population inversion,
and the phase $\pi/2$ ($\pi$) of the final state for $\Psi=\pi/2$ ($\Psi=0$) is the global phase, which can be ignored.
As a result, the transition probability of the state $|f\rangle$ is only associated with its amplitude, and the
Taylor series is
\begin{equation}
P_{2} (\epsilon)=1+O(\epsilon^4),\nonumber\\
\end{equation}
which is accurate to the fourth order in the pulse area error $\epsilon$.

\end{widetext}

\end{appendix}

\bibliographystyle{apsrev4-1}
\bibliography{reference}

\end{document}